\documentclass[prd,nofootinbib,showpacs,preprint]{revtex4}
\usepackage{graphicx}
\usepackage{epsfig}
\usepackage{amsmath}
\usepackage{amsfonts}
\usepackage{amssymb}
\usepackage{url}
\usepackage{hyperref}
\usepackage{subfigure}
\usepackage{cancel}
\newcommand{\beqa}{\begin{eqnarray}}
\newcommand{\eeqa}{\end{eqnarray}}
\newcommand{\beq}{\begin{equation}}
\newcommand{\eeq}{\end{equation}}

\newcommand{\nn}{\nonumber}
\newcommand{\bmt}{\begin{pmatrix}}
\newcommand{\emt}{\end{pmatrix}}
\usepackage[toc,page]{appendix}
\usepackage{comment}
\newcommand{\be}{\begin{equation}}
\newcommand{\ee}{\end{equation}}
\newcommand{\bea}{\begin{eqnarray}}
\newcommand{\eea}{\end{eqnarray}}

\begin{document}
\title{Lepton flavour violating $B$  meson decays via scalar leptoquark}
\author{Suchismita Sahoo and Rukmani Mohanta }
\affiliation{\,School of Physics, University of Hyderabad,
              Hyderabad - 500046, India  }
\begin{abstract}
We study the effect of scalar leptoquarks in the lepton flavour violating $B$ meson decays induced by the flavour changing 
transitions $b \to q l_i^+ l_j^-$ with $q=s,d$. In the standard model these transitions are extremely rare as they are
either two-loop suppressed or proceed via box diagrams with  tiny neutrino masses in  the loop. However, in the leptoquark model they can occur at tree level
and are expected to have  significantly large branching ratios.
The  leptoquark parameter space is constrained using the experimental limits on the branching ratios of $B_q \rightarrow l^+ l^-$ processes.
 Using such constrained   parameter space, we  predict  the branching ratios of LFV semileptonic $B$  meson decays, 
such as $B^+ \to K^+ (\pi^+) l_i^+ l_j^-$, $B^+ \to (K^{* +}, \rho^+) l_i^+ l_j^-$  and $B_s \to \phi l_i^+ l_j^-$,
which are found to be within the experimental reach of LHCb and the upcoming Belle II experiments. We also investigate the rare leptonic $K_{L, S} \to \mu^+ \mu^- (e^+ e^-)$ and $K_{L} \to \mu^\mp e^\pm$ decays in the leptoquark model.
\end{abstract}
\maketitle
\section{Introduction}

The study of rare $B$ decay modes induced by the flavour changing neutral current (FCNC) transitions are immensely helpful to test 
the standard model (SM) and provide hints for new physics beyond it. The SM contributions to the rare $B$ meson decays 
involving  FCNC transitions $b \to s, d$  are absent at the tree level due to  Glashow-Iliopoulos-Maiani (GIM) mechanism 
and occur via one-loop level only. Although, so far we have not observed any  clear indication of new physics in the $B$ sector,
but there are several observables measured by the BaBar, Belle, CDF and LHCb Collaborations in the  semileptonic $B$  decays
involving the transition $b \to s l^+ l^-$,   
have significant deviations from the corresponding SM predictions.
Specifically, the observation of 3$\sigma$ anomaly by the LHCb experiments 
 in the $B \to K^{* 0} \mu^+ \mu^-$  decay rate \cite{lhcb1, lhcb3} and  in the angular  observable $P_5^\prime$ \cite{lhcb2} have 
attracted a lot of attention in recent times. In addition,  lepton flavour 
non-universality has been observed in the ratio of $B \to K \mu \mu $ to $B \to K e e $ branching fractions \cite{lhcb4} and  the decay rate of  
$B_s \to \phi \mu^+ \mu^-$  process \cite{lhcb5} also deviates from the SM predictions by about 3$\sigma$.

On the other hand, the observation of neutrino oscillations has provided unambiguous evidence  for lepton number violation  in the neutral lepton sector, 
even though the individual lepton number is conserved in the SM of the electroweak interaction. The observation of the neutrino masses and mixing
and the violation of family lepton number could in principle allow FCNC transitions  in the charged lepton sector as well,  such as $l_i \to l_j \gamma$, 
$l_i \to l_j l_k \bar{l}_k$, $B \to l_i^\pm l_j^\mp$ and $B \to K^{(*)} l_i^\pm l_j^\mp$ etc. It is interesting to see if these branching ratios could be 
enhanced in some new physics model which could simultaneously explain the observed anomalies. 
 As LHCb has  already reported violation of lepton universality in the $B \to K l^+ l^-$ process having deviation from the SM prediction by 2.6$\sigma$, 
which in turn hints towards the possibility  of observing  lepton flavour violating (LFV) decays also. As pointed out in Ref. \cite{glashow}, 
 a possible explanation for the observed LHCb data on $R_K$, i.e., the lepton non-universality  is due to 25$\%$ deficit in the muon channel, 
which implies LFV is larger 
for muons than for electrons. The LFV decays in the charged lepton sector has been studied in various new physics model in the literature \cite{lfv, mohanta2}. 
Even though there is no direct experimental evidence for such processes, but there exist severe constraints on some of these LFV modes \cite{pdg}. 
The experimental observation of lepton flavour violating decays  would provide unambiguous signal of  new physics  beyond the SM.

The most elegant ways to look for new physics in  FCNC processes are the prudent investigation  of the anomalies associated with 
$b \to s l^+ l^-$ decays observed at LHCb \cite{lhcb1, lhcb2, lhcb3, lhcb4, lhcb5}. These anomalies have been studied  in the SM and 
in various extensions of it \cite{matias1, jager, huber, beaujean}. In this paper we are interested to investigate the lepton 
flavour violating $B$ meson  decays such as $B^+ \to K^+ (\pi^+) l_i^+ l_j^-$,  $B^+ \to (K^{*+}, \rho^+) l_i^+ l_j^-$, 
$B_s \to \phi l_i^+ l_j^-$, $K_{L,S} \to \mu^- \mu^+(e^- e^+)$ and $K_{L} \to \mu^\pm e^\mp$ in the scalar leptoquark model. These decay processes are extremely rare in the SM as they are either
two-loop suppressed or proceed through box diagrams with  the presence of tiny neutrino masses in  the loop. However, in the leptoquark model they can 
occur at the tree level and hence, can give  observable signature in the LHCb experiment.  Leptoquarks are color triplet gauge particles having both
baryon and lepton quantum numbers and can be either scalars or vectors. Even though leptoquarks do not address some important open questions 
like the dark matter content 
of the Universe or the origin of the electroweak scale, these particles allow quark-lepton transitions at tree level and thus,
point towards the theory of  quark-lepton universality.   
 Leptoquarks can come from the extended Standard models \cite{georgi} which treat quarks and leptons in equal footing such as Grand unified model 
\cite{georgi, georgi2}, Pati-Salam model, quark and lepton composite models \cite{kaplan}, extended technicolor model \cite{schrempp} etc. 
Leptoquarks having baryon and lepton number violating couplings are very massive to avoid proton decay or large Majorana neutrino masses. 
However, the baryon and lepton number conserving leptoquarks could be light enough to be accessible in accelerator searches and 
also they do not induce  proton decay. In the literature  there 
are many attempts \cite{mohanta1, mohanta2, mohanta3, davidson, arnold, kosnik, leptoquark, dorsner} to explain the observed anomalies 
in the  leptoquark model.

The outline of the paper is follows. The effective Hamiltonian describing the  $b \rightarrow q l^+ l^-$, $q = s, d$ processes is 
briefly discussed  in section II.  
In section III  we present the new physics contribution due to the scalar leptoquark exchange and  the constraint on leptoquark parameter space 
by using the experimental limit on the  branching ratios of the rare decays  $B_q \rightarrow l^+ l^- $.
The branching ratios for LFV  decays 
$B^+ \to P^+ l_i^+ l_j^-$, $P = K, \pi$ and $B_{(s)}^+ \to V^+(\phi)  l_i^+ l_j^-$, $V =  K^{*}, \rho$ decays in the leptoquark model 
are presented  in  sections IV  and V respectively. In section VI we compute the branching ratios of rare $K_{L, S} \to \mu^+ \mu^- (e^+ e^-)$ 
decays and the LFV decays $K_{L} \to \mu^\mp e^\pm$ are investigated in section VII. Section VIII contains the summary and conclusion.

\section{ The Effective Hamiltonian for $b \to (s,d) l^+ l^-$ process }

 The effective Hamiltonian mediating the rare semileptonic decay $ b \to q l^+ l^-$, $q = s, d$ in the standard model is \cite{Beneke, kohda}
\bea
{\cal H}_{eff} = - \frac{G_F}{\sqrt 2} \left [\lambda_t^{(q)}{\cal H}_{eff}^{(t)}
+ \lambda_u^{(q)}{\cal H}_{eff}^{(u) }\right ]+ h.c., \label{ham-sm}
\eea
where
\bea
{\cal H}_{eff}^{(u)}&=& C_1(\mathcal{O}_1^c-\mathcal{O}_1^u)+C_u(\mathcal{O}_2^c -\mathcal{O}_2^u), \nn\\
{\cal H}_{eff}^{(t)}&=& C_1 \mathcal{O}_1^c + C_2 \mathcal{O}_2^c + \sum_{i=3}^{10} C_i \mathcal{O}_i
\;,
\eea
 and $\lambda_{q^\prime}^{(q)}=V_{q^\prime b}V_{q^\prime q}^*$  ($q^\prime = t,u$) are the product of Cabibbo-Kobayashi-Maskawa (CKM) matrix elements. 
Here $\mathcal{O}_i$'s are the most general dimension-six flavour changing operators and $C_{i = 1,..,10}$ are 
their respective Wilson coefficients evaluated at renormalization scale $\mu = m_b$ \cite{kohda}. The sum over $i$ corresponds to the tree level 
current-current operators ($\mathcal{O}_{1,2}$), the QCD penguin operators $\mathcal{O}_{3-6}$, the photon and gluon dipole operators 
$\mathcal{O}_{7,8}^{(\prime)}$ and the semileptonic operators $\mathcal{O}_{9,10}^{(\prime)}$ which can be expressed as
\bea
\mathcal{O}_{7}^{(\prime)} &=& \frac{e}{16 \pi^2} \left[\bar{s}\sigma_{\mu \nu} (m_s P_{L (R)} +m_b P_{R (L)})b\right] F^{\mu \nu}\;,  \nn \\
\mathcal{O}_{9}^{(\prime)} &=& \frac{\alpha}{4\pi} \left(\bar{s} \gamma^\mu P_{L (R)} b\right) \left(\bar{l}\gamma_\mu l\right), \hspace{1cm} 
\mathcal{O}_{10}^{(\prime)} = \frac{\alpha}{4\pi} \left(\bar{s} \gamma^\mu P_{L (R)} b\right) \left(\bar{l}\gamma_\mu \gamma_5 l\right).
\eea
It should be noted that the  primed operators which have opposite chirality to the unprimed ones  are negligible in the SM and can only be generated using new physics beyond the SM.
The Fermi constant is denoted by  $G_F$, $\alpha$ is the fine-structure constant and $P_{L,R} = (1\mp \gamma_5)/2$ are the chiral operators. 
For $b \to s $ transitions, the contribution of ${\cal H}_{eff}^{(u)}$ is doubly Cabibbo-suppressed with respect to that of ${\cal H}_{eff}^{(t)}$ due to the  
CKM factor $V_{ub} V_{us}^*$ and can be neglected. However,  for $b \to d $ transition, $\lambda_{t}^{(d)}$ and $\lambda_{u}^{(d)}$ 
are comparable in magnitude 
with a sizable phase difference, hence, in addition with ${\cal H}_{eff}^{(t)}$, the decay amplitude from ${\cal H}_{eff}^{(u)}$ is  also relevant. 
In the next section, we will discuss the new physics contribution to the SM effective Hamiltonian due to the exchange of scalar leptoquark and 
constrain the product of various leptoquark couplings from some rare $B$ decays.
\section{New physics contributions due to scalar leptoquark exchange}

There will be additional contributions to the SM effective Hamiltonian (\ref{ham-sm}) in the scalar leptoquark model 
due to the exchange of LQ's between the external fermion particles. As discussed in \cite{arnold,mohanta2},
out of all possible leptoquark multiplets which are invariant under the SM gauge group $SU(3) \times SU(2) \times U(1)$
the two scalar leptoquark multiplets  $X=(3,2,7/6)$ and $X=(3,2,1/6)$ do not allow proton decay. These scalar leptoquarks can have 
sizable Yukawa couplings and could potentially contribute to the quark level transition $b \to q l^+ l^-$. Due to the chirality 
and diagonality nature   and the conservation of both baryon and lepton number,  these leptoquarks may provide an  interesting 
testing ground to look for their effects in rare $B$ meson decays.

In the scalar LQ  model, the Lagrangian describing the  interaction of the scalar leptoquark doublet $X = (3,2,7/6)$ with the charged leptons is given 
by \cite{arnold}
\bea
{\cal L} = - \lambda_{u}^{ij}~ \bar u_R^i X^T \epsilon L_L^j - \lambda_e^{ij}~\bar e_R^i X^\dagger Q_L^j + h.c.\;, \label{lq7-lagrangian}
\eea
where $i,j$ are the generation indices,  $Q_L$ and $L_L$ are the left handed quark and lepton doublets,
$u_R$ and $e_R$ are the right handed up-type quark and charged lepton singlets and $\epsilon=i \sigma_2$ is a $2\times 2$ matrix. 
These multiplets can be represented more explicitly as
\bea
X=
\left( \begin{array}{c}
 V_\alpha   \\
Y_\alpha
\end{array}
\right ), \; \;\:\:Q_L =
\left ( \begin{array}{c}
u_L\\
d_L \\
\end{array}
\right ) ,~~~
L_L =
\left ( \begin{array}{c}
\nu_L\\
e_L \\
\end{array}
\right ) ,~~~~{\rm and}~~~~~
\epsilon=
\left( \begin{array}{cc}
 0 & 1  \\
-1~ & 0 \\
\end{array}
\right ).
\eea
After expanding the $SU(2)$ indices the interaction Lagrangian becomes
\bea
{\cal L}= -\lambda_u^{ij}~ \bar u_{\alpha R}^i ( V_\alpha e_L^j - Y_\alpha \nu_L^j )
-\lambda_e^{ij}~ \bar e_R^i \left (V_\alpha^\dagger u_{\alpha L}^j + Y_\alpha^\dagger d_{\alpha L}^j \right )+h.c.\;.\label{lepto}
\eea
 After performing the  Fierz transformation in (\ref{lepto}) and then comparing it with  the SM effective Hamiltonian (\ref{ham-sm}),  
one can obtain the  new Wilson coefficients  to the  $b \to q l_i^+ l_i^-$ processes as
\bea
C_9^{NP} = C_{10}^{NP} = - \frac{ \pi}{2 \sqrt 2 G_F \alpha V_{tb} V_{tq}^* }\frac{\lambda_e^{i3}{ \lambda_e^{ik}}^*}{
M_Y^2}\;,\label{c10np}
\eea
where $k$ is the generation index of the quark flavor $q$.
Analogously, the Lagrangian for the coupling of the scalar leptoquark $X = (3,2,1/6)$   to the fermion bilinear is
 \bea
{\cal L} = - \lambda_d^{ij}~ \bar{d}_{\alpha R}^i (V_\alpha e_L^j-Y_\alpha \nu_L^j) +h.c.\;,
\eea
which provides additional contributions to the primed semileptonic electroweak penguin operators  $\mathcal{O}_{9,10}^{\prime}$ 
and their corresponding  new  primed Wilson coefficients $C^{\prime NP}_{9,10}$ are given as
\bea
C_9^{'NP } = - C_{10}^{'NP } = \frac{ \pi}{2 \sqrt 2 ~G_F \alpha V_{tb}V_{tq}^*} \frac{\lambda_d^{ki} {\lambda_d^{3i}}^*}{M_V^2}\;.\label{c10np1}
\eea
Here the superscript index $k$ represents the generation of the down type quark, i.e., $k = 1 ~ {\rm or}~ 2$, depending upon the coupling of leptoquark 
to  $ d$  or $s$.
The subscript $d(e)$ in the leptoquark couplings of Eqns. (\ref{c10np}) and (\ref{c10np1}) stands for all families of  down-type quarks (charged leptons).
These new Wilson coefficients $C_{9,10}^{(\prime) NP}$ or in other words, the leptoquark parameters can be constrained  by comparing the theoretical \cite{bobeth1} 
and experimental \cite{cms, lhcb6, lhcb7}  branching ratios of $B_q \to \mu^+ \mu^-$   processes. The detailed formalism 
of the constraints on leptoquark coupling  has been discussed in \cite{mohanta1, mohanta2}, therefore here we will simply quote  the  results.  
\bea
 1.5 \times 10^{-9} ~{\rm GeV^{-2}} \leq \frac{|\lambda_e^{23} {\lambda_e^{21}}^*|}{M_S^2}=\frac{|{\lambda_d^{32}}^* {\lambda_d^{12}}|}{M_S^2} \leq 3.9 \times 10^{-9}~ {\rm GeV^{-2}}\;,\hspace*{1.5 true cm} 
 \eea
\bea
 0 \leq \frac{|\lambda_e^{23} {\lambda_e^{22}}^*|}{M_S^2}=\frac{|{\lambda_d^{32}}^* {\lambda_d^{22}}|}{M_S^2} \leq 5 \times 10^{-9}~ {\rm GeV^{-2}} ~~~~{\rm for}~~~~\pi/2 \leq \phi^{NP} \leq 3 \pi/2\;,
 \eea
where $M_S$ is the mass of the scalar leptoquark. Also for simplicity, we have not kept the subscripts on the leptoquark coupling parameters.
Analogously from   the $B_{s,d} \to \tau^+ \tau^- ~(e^+ e^-)$ leptonic decays, the constraints on  
various combination of leptoquark couplings can be obtained by comparing the 
 theoretically predicted branching ratios \cite{bobeth1} 
with the corresponding experimental ones, for which only the upper limits  are known \cite{pdg}. The upper bound on the product of the 
leptoquark couplings from various two body leptonic $B_{s,d} \to l^+ l^-$, $l = e, \mu, \tau$ decays are presented in Table-I \cite{mohanta2}.
 In our previous work \cite{mohanta2, mohanta3}, we studied the  effect of scalar letoquarks on various observables associated with 
$B \to K^{(*)} \mu^+ \mu^- (\nu \bar \nu)$ and $B^+ \to \pi^+ \mu^+ \mu^-$ processes  by using these constraint leptoquark couplings. 
We found significant deviation in the asymmetry parameters from their SM predictions and thus explains the anomalies observed at LHCb and other 
$B$-factories quite well.

\begin{table}[htb]
\begin{center}
\caption{Constraints obtained from the leptoquark couplings from various leptonic $B_{s,d} \to l^+ l^-$ decays.}
\vspace*{0.1 true in}
\begin{tabular}{|c|c|c|}
\hline
Decay Process ~& ~Couplings involved ~&~ Upper bound of  \\
             &  &~the couplings (${\rm GeV^{-2}}$)~  \\
\hline
$B_s \to \mu^\pm \mu^\mp $~~ &~~ $\frac{|\lambda^{23} {\lambda^{22}}^*|}{M_S^2}$ ~~& ~~$ \leq 5 \times 10^{-9} $~\\

\hline

$B_s \to e^\pm e^\mp $ &~ $\frac{|\lambda^{13} {\lambda^{12}}^*|}{M_S^2}$ ~& ~$ < 2.54 \times 10^{-5} $~\\

\hline

$B_s \to \tau^\pm \tau^\mp $ &~ $\frac{|\lambda^{33} {\lambda^{32}}^*|}{M_S^2}$ ~& ~$ < 1.2 \times 10^{-8} $~\\

\hline
$B_d \to \mu^\pm \mu^\mp $ &~ $\frac{|\lambda^{23} {\lambda^{21}}^*|}{M_S^2}$ ~& ~$ (1.5 -3.9 ) \times 10^{-9} $~\\

\hline
$B_d \to e^\pm e^\mp $ &~ $\frac{|\lambda^{13} {\lambda^{11}}^*|}{M_S^2}$ ~& ~$ < 1.73 \times 10^{-5} $~\\

\hline

$B_d \to \tau^\pm \tau^\mp $ &~ $\frac{|\lambda^{33} {\lambda^{31}}^*|}{M_S^2}$ ~& ~$ < 1.28 \times 10^{-6} $~\\

\hline
\end{tabular}
\end{center}
\end{table}

\section{$B^+ \rightarrow P^+ l^+_i  l^-_j$}

Here, we will discuss the lepton flavour violating semileptonic $B$ mesons decays to pseudoscalar mesons $K$ and $\pi$, which are  mediated by  the 
$b \to q l_i^+ l_j^-$ quark level transition. As discussed earlier, these processes occur at tree level due to the exchange of scalar
leptoquarks. Fig. 1 depicts the tree level Feynman diagram for the lepton flavour violating process $b \to s l_i^+ l_j^-$, 
where leptoquark can couple to a quark and a lepton simultaneously. Analogously, one can obtain the diagram for $b \to d l_i^+ l_j^-$ process
by replacing $s$ with $d$ and incorporating the appropriate LQ couplings.  Here $i,j$ denote the lepton family numbers. 
We will present  the  results for the scalar LQ $X(3,2,7/6)$ and analogously one can obtain the results for $X(3,2,1/6)$. 
Thus, the effective Hamiltonian for $b \to q l_i^\pm l_j^\mp$ process in the  scalar LQ  model is given by \cite{mohanta2}
\bea
{\cal{H}}_{LQ} = \left[ G_{LQ} \left( \bar{q}\gamma^\mu P_L b \right) (\bar{l}_i\gamma_\mu (1+ \gamma_5) l_j) + 
 H_{LQ} \left( \bar{q}\gamma^\mu P_L b \right) (\bar{l}_j\gamma_\mu (1+ \gamma_5) l_i) \right]\;,\label{ham-lq}
\eea
where the constant coefficient $G_{LQ}$ and $H_{LQ}$ are
\be
G_{LQ} = \frac{\lambda^{i3} {\lambda^{jk}}^*}{8M_Y^2}, \hspace{1cm}  H_{LQ} = \frac{\lambda^{j3} {\lambda^{ik}}^*}{8M_Y^2} \;.
\ee
\begin{figure}[h]
\centering
\includegraphics[scale=0.55]{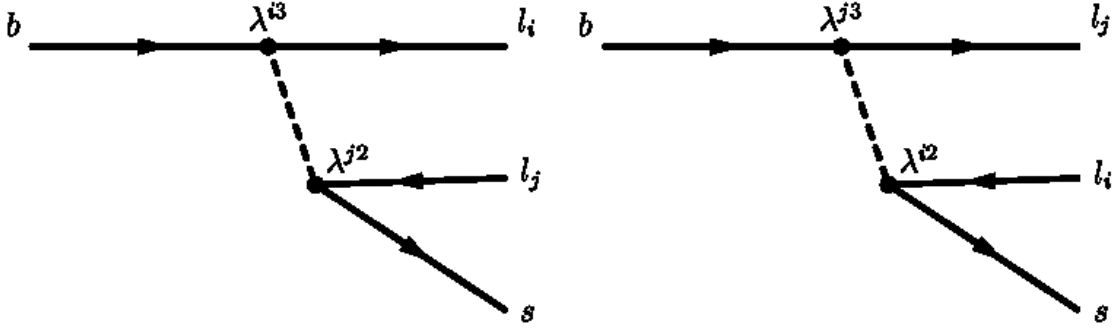}
\caption{Feynman diagram for lepton flavour violating $b \to s l_i^- l_j^+$ process (left panel) and
$b \to s l_i^+ l_j^-$ (right panel) mediated by the scalar leptoquark where  $l = e, \mu, \tau$.}
\end{figure}

The matrix element of the quark-current between the initial and final mesons   can be parameterized in terms of
the form factors  $f_+^P$ and  $f_0^P$ as \cite{bobeth2}
\bea
\langle P(p^\prime)| \bar s \gamma_\mu b | \bar{B}(p_B) \rangle =(2 p_B - q)_\mu f_+^P(q^2) + \frac{M_B^2-M_P^2}{q^2}q_\mu [f_0^P (q^2)-f_+^P (q^2)]\;,
\eea
where $q = p_B - p^\prime$ and $f_{+, 0}^P$ correspond to  kaon and pion form factors, which are taken from \cite{kaonformfactor} 
and \cite{piformfactor} respectively.
The  transition amplitudes for $B^+ \to P^+ l_i^- l_j^+$, $P = K, \pi$ processes are  given as
\begin{equation}
\mathcal{M} =  \Bigg[F_S (\bar{l}_i l_j) + F_P (\bar{l}_i \gamma_5 l_j) + F_V P^\mu \left(\bar{l}_i \gamma_\mu l_j\right) + F_A P^\mu \left(\bar{l}_i \gamma_\mu 
\gamma_5 l_j\right) \Bigg]\;,\label{eq:13}
\end{equation}
where
\bea
F_V &=& G_{LQ}\;, \hspace{3cm} F_A = G_{LQ}\;, \nn\\
F_S &= & \frac{1}{2} G_{LQ} (m_j - m_i) \Bigg[\frac{M_B^2 - M_P^2}{q^2}  \left( \frac{f_0^P (q^2)}{f_+^P (q^2)} - 1 \right) -1 \Bigg]\;,\nn\\
F_P & = & \frac{1}{2} G_{LQ} (m_i + m_j)\left[\frac{M_B^2 - M_P^2}{q^2} \left( \frac{f_0^P (q^2)}{f_+^P (q^2)} - 1 \right)-1 \right]\;.\label{amp}
\eea
Analogously, the transition amplitude for  $B^+ \to P^+ l_i^+ l_j^-$ process can be obtained from (\ref{eq:13}) by replacing $G_{LQ}$ by $H_{LQ}$
and $l_i \leftrightarrow l_j$.

Thus, one can obtain the differential decay distribution for the process $B^+ \to P^+ l_i^+ l_j^-$, with respect to $q^2$ and $\cos\theta$ as
\begin{equation}
\frac{d\Gamma}{dq^2 d\cos\theta} = a (q^2) + b (q^2) \cos\theta + c (q^2) \cos^2\theta\;,
\end{equation}
where
\bea
a (q^2) & =& \Gamma_0 \frac{\sqrt{\lambda_1 \lambda_2}}{q^2} (f_+^P)^2 \Bigg[ \left(|F_V|^2 + |F_A|^2 \right) \frac{\lambda_1}{4}
 + |F_S|^2 \left (q^2 - (m_i + m_j)^2\right )\nn\\ 
& + & |F_P|^2 \left (q^2 - (m_i - m_j)^2 \right )  + |F_A|^2 M_B^2 (m_i + m_j)^2 + |F_V|^2 M_B^2 (m_i - m_j)^2\nn  \\
& + & \left(M_B^2 - M_P^2 + q^2 \right) \Big( (m_i + m_j)  Re (F_P F_A^*) + (m_j - m_i)  Re (F_S F_V^*)\Big) \Bigg]\;,
\eea
\begin{equation}
b (q^2) = \Gamma_0 \frac{\sqrt{\lambda_1 \lambda_2}}{q^2} (f_+^P)^2 \Bigg[ (m_i + m_j) Re (F_S F_V^*) - (m_j - m_i) Re (F_P F_A^*) \Bigg]\;, \hspace{1.2cm}
\end{equation}
\begin{equation}
c (q^2) = - \Gamma_0  (f_+^P)^2 \frac{(\lambda_1 \lambda_2)^{3/2}}{4 q^6} \left(|F_A|^2 + |F_V|^2 \right)\;, \hspace{6cm}
\end{equation}
and
\bea
\Gamma_0  =  \frac{1}{2^8 \pi^3 M_B^3}, \hspace{1cm} \lambda_1 = \lambda (M_B^2, M_P^2, q^2),  \hspace{1cm} \lambda_2 = \lambda (q^2,m_i^2, m_j^2),  \nn \\
{\rm with}~~~~\lambda(a,b,c) = a^2+b^2+c^2-2(ab+bc+ac)\;. \hspace{3cm}
\eea
It should be noted that,  in the SM there are no  intermediate states which can decay into two leptons belonging to different generations. 
Therefore, LFV decays have no long distance 
QCD contributions and no dominant charmonium resonance background like $B \to K^{(*)} l^+ l^-$ processes. Therefore, the
background suppression for these channels would be relatively low.

After obtaining the expression for the branching ratio of $B^+ \to P^+ l_i^+ l_j^- $ processes, we will proceed 
for numerical estimations. The particle masses and the life time of $B$ meson are taken from \cite{pdg}. The  scalar leptoquarks are diagonal 
and  have full strength coupling when couples to a lepton and a quark of the same generation. The coupling of LQ with quark and leptons of
different generations are assumed to follow
Cabibbo-like suppression behavior. It should be noted that   the expansion parameter of the CKM matrix in the Wolfenstein parameterization 
can be related to the down type quark masses as $\lambda \sim (m_d/m_s)^{1/2}$, while in the lepton sector one can have the same order 
for $\lambda$ with the relation $\lambda \sim (m_{l_i}/m_{l_j})^{1/4}$. Therefore, in order to compute the  required couplings, we used the 
coupling given in Table-I as basis values and  assumed that the leptoquark  couplings between different
generation of quarks and leptons follow the simple scaling law, i.e. $\lambda^{ij} \simeq (m_i/m_j)^{1/4}\lambda^{ii}$ with $j \textgreater i$.

With these input parameters we show in Fig. 2
 the variation of  branching ratio  for lepton flavour violating decays $B^+ \to K^+ \mu^+ e^-$ (left panel), $B^+ \to K^+ \tau^+ e^-$ (right panel) 
and $B^+ \to K^+ \tau^+ \mu^-$ (bottom panel) with respect to $q^2$ in the full physical region.  In Fig. 3, we have shown the variation 
of branching ratios of $B^+ \to \pi^+ \mu^+ e^-$ (left panel), $B^+ \to \pi^+ \tau^+ e^- $ (right panel) and $B^+ \to \pi^+ \tau^+ \mu^- $ (bottom panel)  processes
with respect to $q^2$. The blue bands represent the allowed range of the branching ratio of semileptonic LFV decays 
$B^+ \to \pi^+ \mu^+ e^-(\tau^+ \mu^-)$ induced by the scalar leptoquarks, as in these cases we have also  the lower bound
on the leptoquark couplings as seen from Table-I. 
The predicted branching ratio of $B^+ \to K^+ (\pi^+) l_i^+ l_j^-$ LFV decays in respective physical range and their corresponding experimental upper 
limits are presented in Table-II.  The predicted branching ratios are  found to be lower than the present experimental upper limits and they are within
the reach of LHCb and Belle II experiments. 

\begin{figure}[h]
\centering
\includegraphics[width=6.0cm,height=5.0cm]{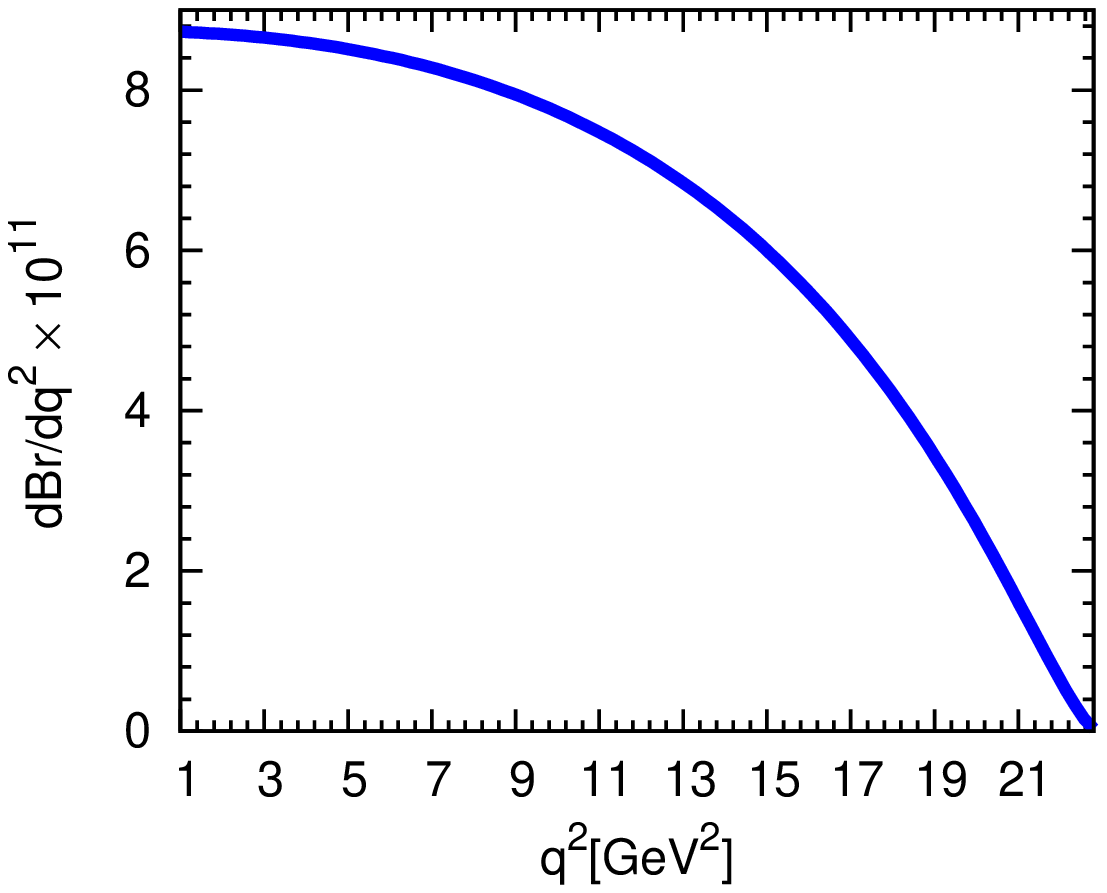}
\includegraphics[width=6.5cm,height=5.0cm]{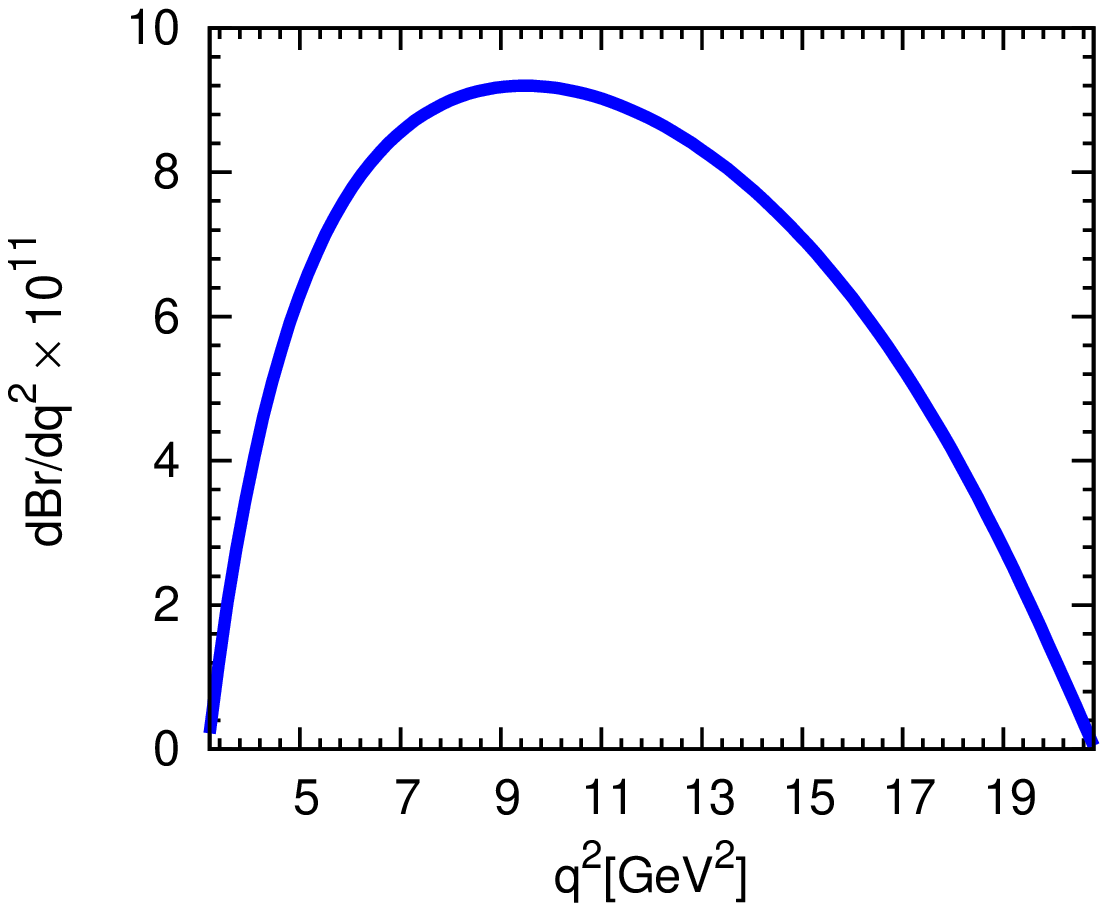}\\
\includegraphics[width=6.5cm,height=5.0cm]{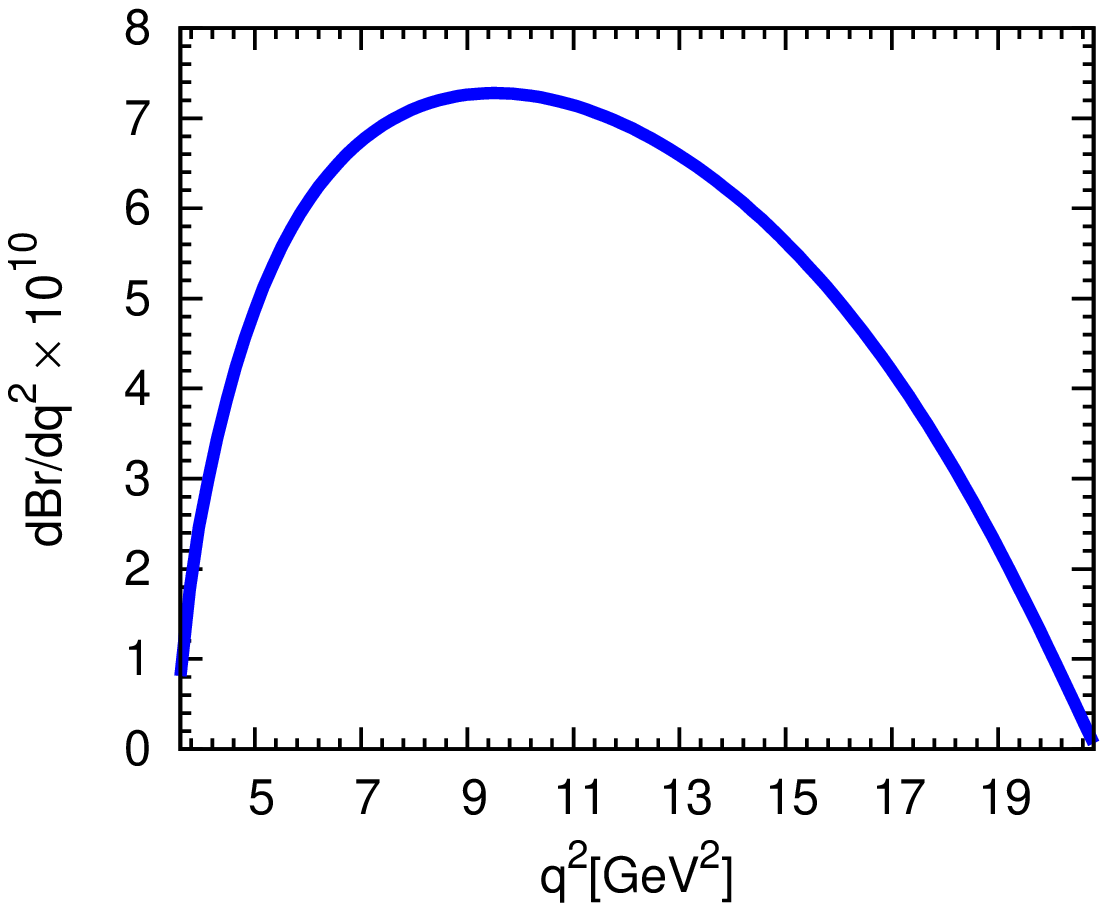}
\caption{The variation of branching ratio of $B^+ \rightarrow K^+ \mu^+ e^-$ (left panel), $B^+ \rightarrow K^+ \tau^+ e^-$ (right panel), and $B^+ \rightarrow K^+ \tau^+ \mu^-$ (bottom panel)  with respect to $q^2$ in the scalar leptoquark model.}
\end{figure}
\begin{figure}[h]
\centering
\includegraphics[width=6.0cm,height=5.0cm]{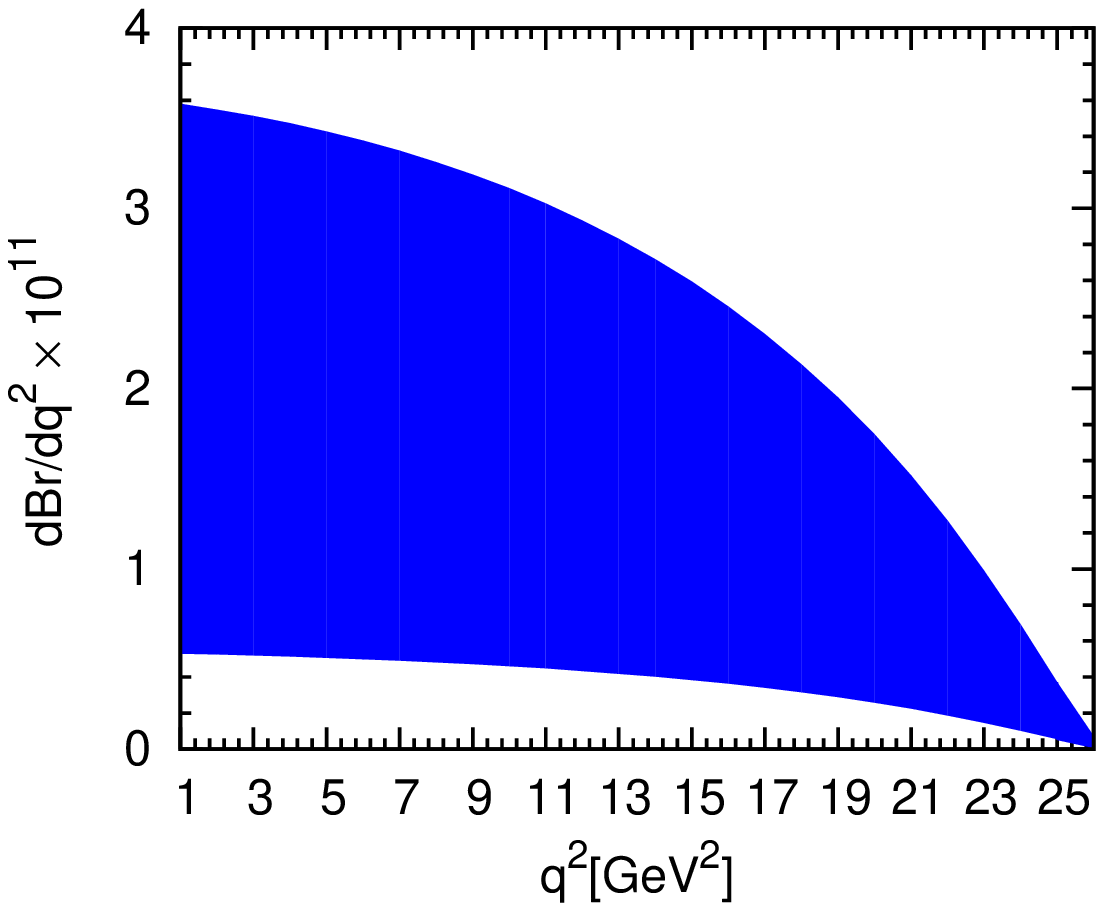}
\includegraphics[width=6.5cm,height=5.0cm]{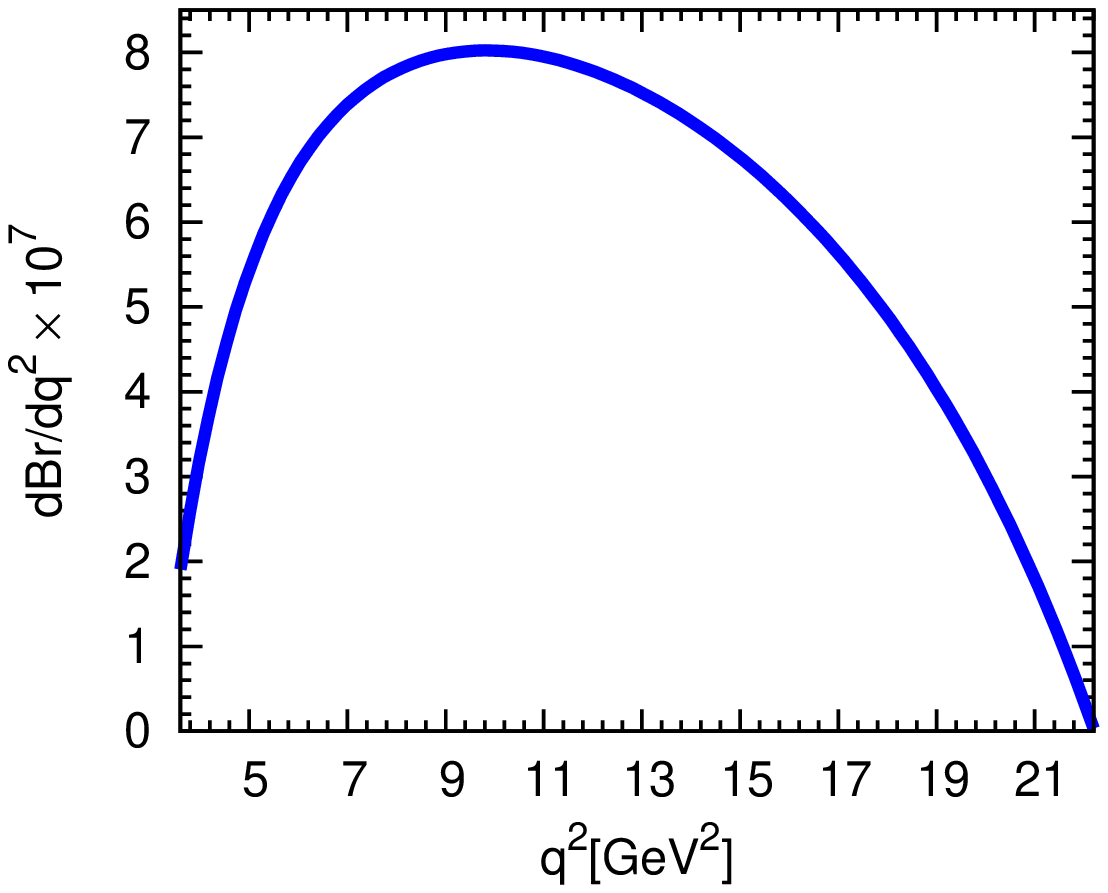}\\
\includegraphics[width=6.5cm,height=5.0cm]{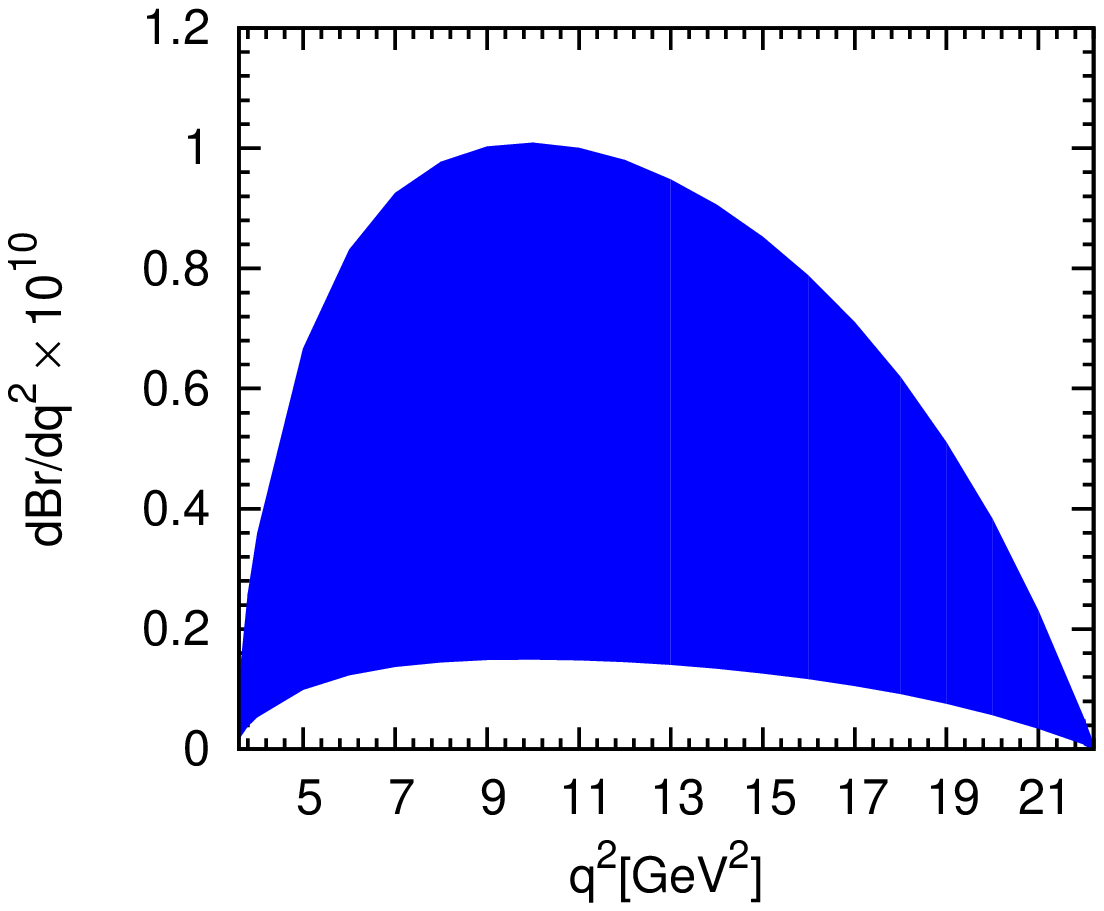}
\caption{The variation of branching ratio of $B^+ \rightarrow \pi^+ \mu^+ e^-$ (left panel), $B^+ \rightarrow \pi^+ \tau^+ e^-$ (right panel), and $B^+ \rightarrow \pi^+ \tau^+ \mu^-$ (bottom panel)  with respect to $q^2$ in the scalar leptoquark model.}
\end{figure}

\begin{table}[h]
\caption{The predicted branching ratios for $B^+ \to K^+ (\pi^+) l_i^+ l_j^-$  lepton flavour violating decays, where $l = e, \mu, \tau$ 
in the scalar LQ $X(3,2,7/6)$ model. }
\begin{center}
\begin{tabular}{| c | c | c |}
\hline
 Decay process & Predicted BR &  Experimental limit \cite{pdg} \\
\hline
$B^+ \to K^+ \mu^+ e^-$ &   $\textless 1.36 \times 10^{-9}$ & $\textless 1.3 \times 10^{-7}$\\
$B^+ \to K^+ \tau^+ \mu^-$ & $\textless 8.8 \times 10^{-9}$ & $\textless 2.8 \times 10^{-5}$\\
$B^+ \to K^+ \tau^+ e^-$ & $\textless 1.12 \times 10^{-9}$ & $\textless 1.5 \times 10^{-5}$\\
\hline
$B^+ \to \pi^+ \mu^+ e^-$ & $(0.91 - 6.16) \times 10^{-10}$ & $\textless 6.4 \times 10^{-3}$\\
$B^+ \to \pi^+  \tau^+ \mu^-$ &$ (0.18 - 1.2) \times 10^{-9}$ & $\textless 4.5 \times 10^{-5}$\\
$B^+ \to \pi^+ \tau^+ e^-$& $\textless  9.65 \times 10^{-6}$ & $\textless 2.0 \times 10^{-5}$\\
 \hline
\end{tabular}
\end{center}
\end{table}
\section{$ B^+ \rightarrow V^+ l^+_i  l^-_j$ and $ B_s \rightarrow \phi l^+_i  l^-_j$}

In this section we describe the theoretical framework to calculate the  branching ratio  for the LFV decays  $B_{(s)}^+ \to V^+ (\phi) l_i^- l_j^+$ , where the vector 
meson  $V$ corresponds to $K^*/\rho$.  Here we will discuss  in detail  for a particular 
vector boson, i.e., $V=K^*$ case. However, the same formalism can be applied to other vector mesons with appropriate change
in the CKM  elements and the  mass of the particles involved. The amplitude of $B(p) \to K^*(k) [\to K(k_1) \pi (k_2)] l_i^+ (p_i) l_j^- (p_j)$ 
decay mediated via the scalar leptoquark can be obtained from the effective Hamiltonian (\ref{ham-lq}) and is given by
\bea
\mathcal{M} = G_{LQ}  \langle K \pi| \bar{s}\gamma _\mu (1-\gamma_5) b | B\rangle   (\bar{l}_i \gamma^\mu (1+\gamma_5) l_j)\;.
\eea
This amplitude can be expressed in terms of $B \to K^*$ form factors by assuming that $K^*$ decays resonantly. 
The $B \to K^*$ hadronic matrix elements of the local quark bilinear operators can be parametrized as \cite{ball1}
\bea
 &&  \langle K^* \left(k \right)| \bar{s}\gamma _\mu (1-\gamma_5) b | B\left(p\right)\rangle =    \epsilon_{\mu \nu \alpha \beta} \epsilon^{*\nu } p^\alpha 
q^\beta \frac{2V(q^2)}{M_B + M_{K^*}} - i \epsilon^*_\mu (M_B + M_{K^*}) A_1(q^2)\nn \\ 
&& \hspace*{2 cm} + i (\epsilon^* \cdotp q)(2p-q)_\mu \frac{A_2(q^2)}{M_B + M_{K^*}} + 
i \frac{2M_{K^*}}{q^2} (\epsilon^* \cdotp q) \left[ A_3(q^2) - A_0(q^2)\right] q_\mu\;,\label{kstar}
  \eea
where $q^2$ is the momentum transfer between the  $B$ and $K^*$ meson, i.e., $q_\mu = p_\mu - k_\mu$ and $\epsilon_\mu$ is the polarization vector of the 
$K^* $ meson. In the narrow width approximation the squared $K^*$ propagator can be expressed as
\bea
\frac{1}{(k^2-M_{K^*}^2)^2 +(M_{K^*} \Gamma_{K^*})^2}~ {\stackrel{\Gamma_{K^*} \ll M_{K^*}}{\xrightarrow{\hspace*{1.5cm}}}}~\frac{\pi}{M_{K^*} \Gamma_{K^*}}
\delta(k^2 -M_{K^*}^2)\;.
\eea 
One can avoid the $K^* K \pi$ coupling $g_{K^* K \pi}$ in the $B \to K \pi$ amplitude as it cancels with the vertex factor and the width
of $K^*$ meson
\be
\Gamma_{K^*} = \frac{g_{K^* K \pi}^2}{48 \pi} M_{K^*} \beta^3\;,
\ee 
where
\bea
 \beta = \frac{1}{M_{K^*}^2} \left[ M_{K^*}^4 + M_K^4 +M_\pi^4 - 2 \left(M_{K^*}^2 M_K^2 +M_K^2 M_{\pi}^2  + M_{K^*}^2 M_{\pi}^2 \right) \right]^{1/2}.
 \eea
If one writes symbolically the $B \to K^* $ matrix elements (\ref{kstar}) as
\be
\langle K^*(k)| \bar{s}\gamma _\mu (1-\gamma_5) b | B\left(p\right)\rangle =\epsilon^{*\nu} A_{\nu \mu}\;, 
\ee
where $A_{\nu \mu}$ contains the $B \to K^*$ form factors, then  $B \to K \pi$ matrix  element can be expressed as
 \bea
 \langle K \pi| \bar{s}\gamma _\mu (1-\gamma_5) b | B\rangle  = -D_{K^ *} (k^2) \left[ K^\nu - \frac{M_K^2 - M_\pi^2}{k^2} k^\nu \right] A_{\nu \mu},
 \eea
 with
 \bea
 |D_{K^*}(k^2)|^2 = g_{K^* K \pi}^2 \frac{\pi}{M_{K^*} \Gamma_{K^*}} \delta (k^2 - M_{K^*}^2) = \frac{48 \pi^2}{\beta^3 M_{K^*}^2}  \delta (k^2 - M_{K^*}^2)\;.
 \eea
  In our analysis, we have used the following symmetric and antisymmetric combination of momentum as
 \bea
 k = k_1 + k_2, \hspace{0.7cm} K = k_1 - k_2, \hspace{0.7cm} q = p_i + p_j, \hspace{0.7cm} Q = p_i - p_j\;.
 \eea

The full angular distribution of $ B \to K^* l_i^+ l_j^-$ decay can be completely described by the four independent kinematic variables, the dilepton invariant mass squared $q^2$, the angle $\phi$ between the normals to the  $K \pi$ and the dilepton $(l_i^+ l_j^-)$  planes in the rest frame of the $B$ meson and the angles $\theta_K$ and $\theta_l$. The physical region of the phase space  are
\bea
 (m_i+m_j)^2 \leq q^2 \leq (M_B-M_{K^*})^2, \hspace{0.25cm} -1 \leq \cos\theta_l \leq 1, \hspace{0.25 cm}
 -1 \leq \cos\theta_K \leq 1, 
\hspace{0.25cm} 0 \leq \phi \leq 2\pi.
\eea
If we integrate out all the three angles $\theta_K$, $\theta_l$ and $\phi$ in their respective kinematically accessible  physical range, 
we will get the the differential decay rate with respect to the dilepton mass squared $q^2$, which is given by
\bea
\frac{d\Gamma}{dq^2} & = & \Gamma_V \times \Bigg[ A(q^2)^2 \Bigg\{ \frac{2}{3} \lambda_{K^*} \left(1-\left( \frac{m_i^2}{q^2} \right)^2\right) 
+  8M_{K^*}^2 (q^2 - m_i^2) \nn \\ && -\frac{2}{9} \left(1-\frac{m_i^2}{q^2}  \right)^2 \left( \left(M_B^2-M_{K^*}^2-q^2\right)^2 
+ 8q^2 M_{K^*}^2 \right) \Bigg\} \nn \\ && + B(q^2)^2 \Bigg\{\frac{\lambda_{K^*}}{6}\left(M_B^2-M_{K^*}^2-q^2\right)^2 \left(1-
\left( \frac{m_i^2}{q^2} \right)^2\right) -\frac{\lambda_{K^*}^2}{18}\left(1-\frac{m_i^2}{q^2}  \right)^2 \nn \\ &&
 -\frac{2}{3}\lambda_{K^*}M_{K^*}^2 (q^2 - m_i^2) \Bigg\}  +C(q^2)^2 \Bigg\{ \frac{2}{3} \lambda_{K^*}m_i^2(q^2 - m_i^2) \Bigg\}\nn \\
 && -D(q^2)^2 \Bigg\{ \frac{4}{9}\lambda_{K^*}M_{K^*}^2(q^2 - m_i^2)\left(4-\frac{m_i^2}{q^2}  \right) \Bigg\} \nn \\ && 
-Re\Big (A(q^2)B(q^2)^*\Big )\Bigg\{ \frac{2}{3}\lambda_{K^*}\left(M_B^2-M_{K^*}^2-q^2\right)\left(1-\left( \frac{m_i^2}{q^2} \right)^2
-\frac{1}{3}\left(1-\frac{m_i^2}{q^2}  \right)^2\right) \Bigg\}  \nn \\ 
&& -Re\Big(A(q^2)C(q^2)^*\Big) \Bigg\{\frac{4}{3}\lambda_{K^*}m_i^2\left(1-\frac{m_i^2}{q^2} \right)\Bigg\}  \nn \\
 && +Re\Big(B(q^2)C(q^2)^*\Big)\Bigg\{ \frac{2}{3}\lambda_{K^*}m_i^2\left(M_B^2-M_{K^*}^2-q^2\right)\left(1-\frac{m_i^2}{q^2} \right)\Bigg\} \Bigg],
\eea
where
\bea
\Gamma_V = \frac{3\sqrt{\lambda_{K^*}}}{2^{11}M_{K^*}^2\left(\pi M_B \beta \right)^3} |G_{LQ}|^2\;, \hspace{1cm} 
\lambda_{K^*} = \lambda \left(M_B^2, M_{K^*}^2, q^2\right)\;,
\eea
and
\bea
A(q^2)&=&\left(M_B+M_{K^*}\right)A_1(q^2)\;, \hspace{2.8cm} B(q^2)=\frac{2A_2(q^2)}{\left(M_B+M_{K^*}\right)}\;, \nn \\  
C(q^2)&=&\frac{A_2(q^2)}{\left(M_B+M_{K^*}\right)}+\frac{2M_{K^*}}{q^2}\left(A_3(q^2)-A_0(q^2)\right)\;, 
\hspace{1cm} D(q^2)=\frac{2V(q^2)}{\left(M_B+M_{K^*}\right)}\;.
\eea
For simplicity, we have neglected the mass of kaon, pion, muon and electron. 
Here $m_i$ represents  the mass of the tau lepton for the  LFV decays having tau as a final particle.
The form factors $A_i$ and $V$ are scale independent and are  taken from \cite{ball2}, which are valid in the full physical regime.
Using the above expressions,  the variation of differential branching ratios of $B^+ \to K^{* +} \mu^+ e^-$ (left panel), 
$B^+ \to K^{* +} \tau^+ e^-$ (right panel) and $B^+ \to K^{* +} \tau^+ \mu^-$ (bottom panel) processes with respect to $q^2$ are shown in Fig. 4.
The predicted branching ratios of the $B^+ \to K^{* +} l_i^+ l_j^-$ processes in the full kinematical regime 
 are presented  in Table-III. It is found that    
the obtained value of  $B^+ \to K^{* +} \mu^+ e^-$ branching ratio is within the experimental limit. 
But so far there exist no experimental upper limits for the  ${\rm Br}(B^+ \to K^{* +} \tau^+ \mu(e)^-)$ decay processes.

\begin{figure}[h]
\centering
\includegraphics[width=6.0cm,height=5.0cm]{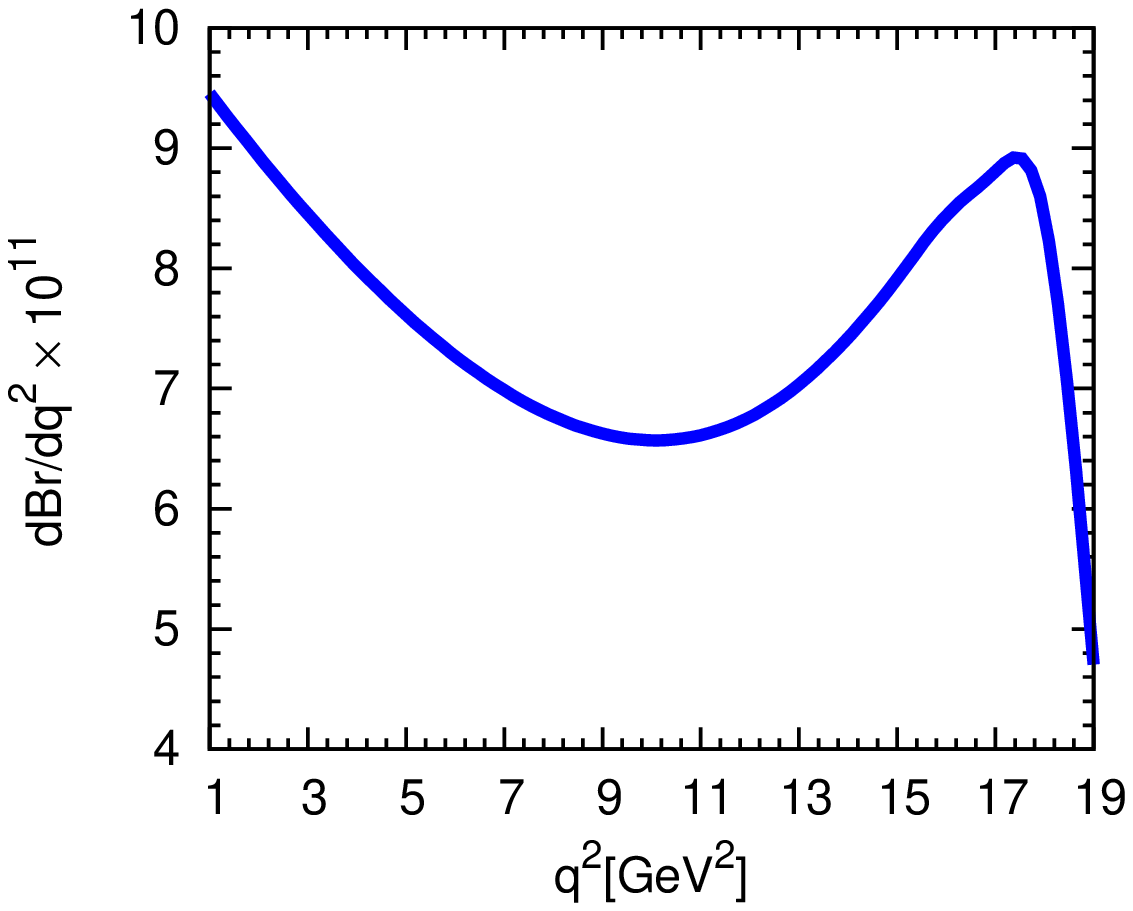}
\includegraphics[width=6.5cm,height=5.0cm]{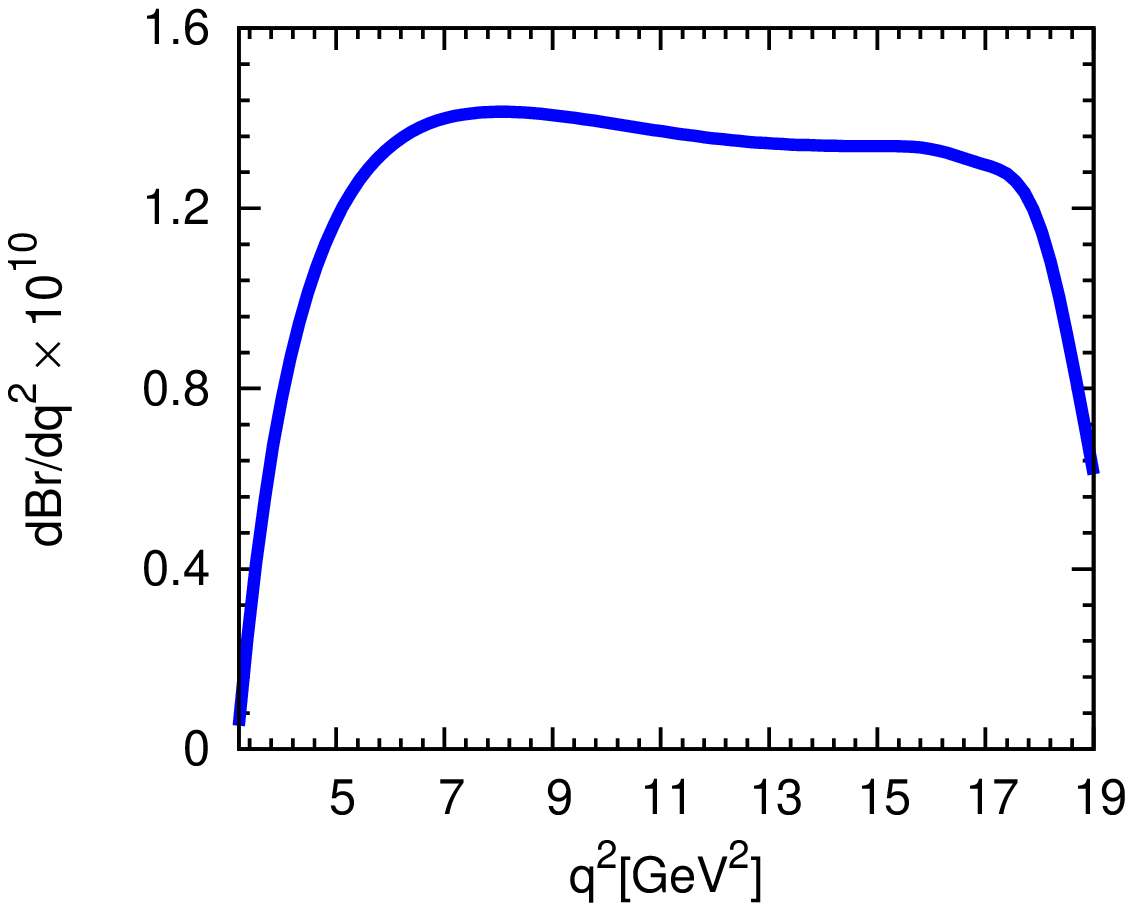}\\
\includegraphics [width=6.5cm,height=5.0cm]{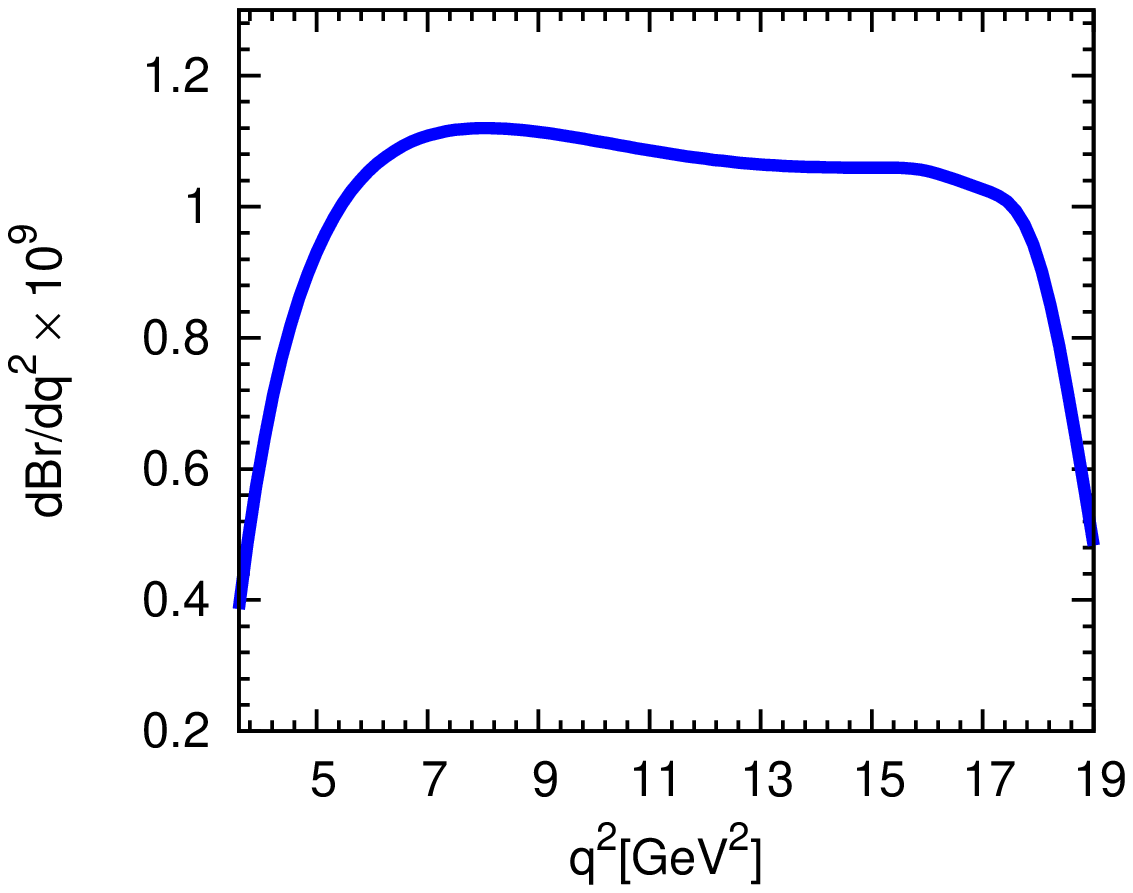}
\caption{The variation of branching ratio of $ B^+ \rightarrow K^{*+} \mu^+ e^-$ (left panel), $B^+ \rightarrow K^{*+} \tau^+ e^-$ (right panel), 
and $ B^+ \rightarrow K^{*+}\tau^+ \mu^-$ (bottom panel)  with respect to $q^2$ in the scalar leptoquark model.}
\end{figure}

Similarly we have estimated branching ratios of the  $B_s \to \phi l_i^+ l_j^-$ and $B^+ \to \rho^+ l_i^+ l_j^-$ processes. 
For numerical calculation, we have taken the values of form factors of $B_q \to \phi(\rho)$ from \cite{ball2} and the particle masses from \cite{pdg}. 
Fig. 5 shows the differential branching ratios of $B_s \to \phi \mu^+ e^-$ (left panel), $B_s \to \phi \tau^+ e^-$ (right panel) and 
$B_s \to \phi \tau^+ \mu^-$ (bottom panel)  
 processes with respect to dilepton invariant mass squared and the branching ratios of $B^+ \to \rho^+ \mu^+ e^-$ (left panel), 
$B^+ \to \rho^+ \tau^+ e^-$ (right panel) and $B^+ \to \rho^+ \tau^+ \mu^-$ (bottom panel) with $q^2$ are shown in Fig. 6. 
The integrated branching ratios of the $B_s \to \phi  l_i^+ l_j^-$ process in the range $(m_i+m_j)^2$ to  $(M_{B_s}-M_\phi)^2 $ are presented
 in Table-III.  Similarly the predicted branching ratio of $B^+ \to \rho^+  l_i^+ l_j^-$ up to full range $(M_B-M_\rho)^2 \simeq 
20.2 ~{\rm GeV}^2$ has been presented in Table-III. The experimental upper limit of $ B^+ \to \rho^+ e^\pm \mu^\mp  $ process is $\textless ~3.2 \times 10^{-6}$ \cite{lfvexpt}.
\begin{figure}[h]
\centering
\includegraphics[width=6.0cm,height=5.0cm]{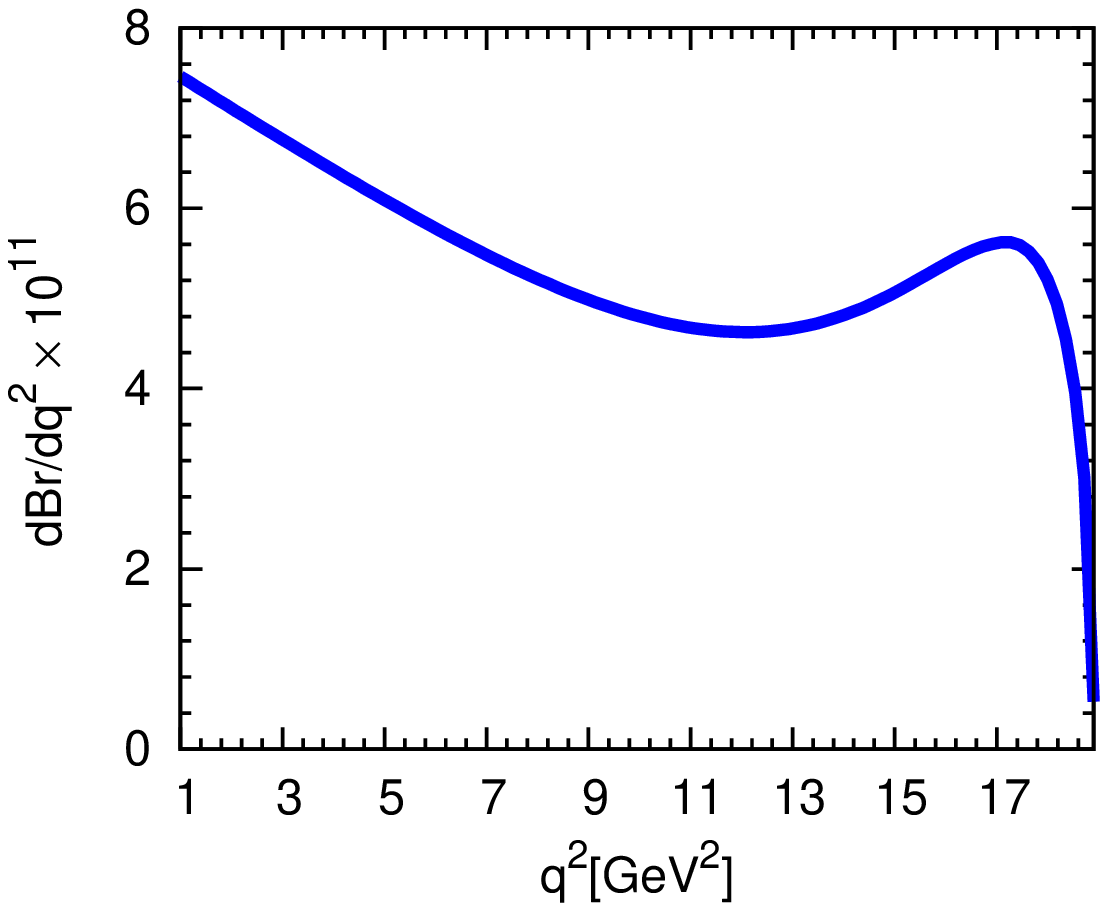}
\includegraphics[width=6.5cm,height=5.0cm]{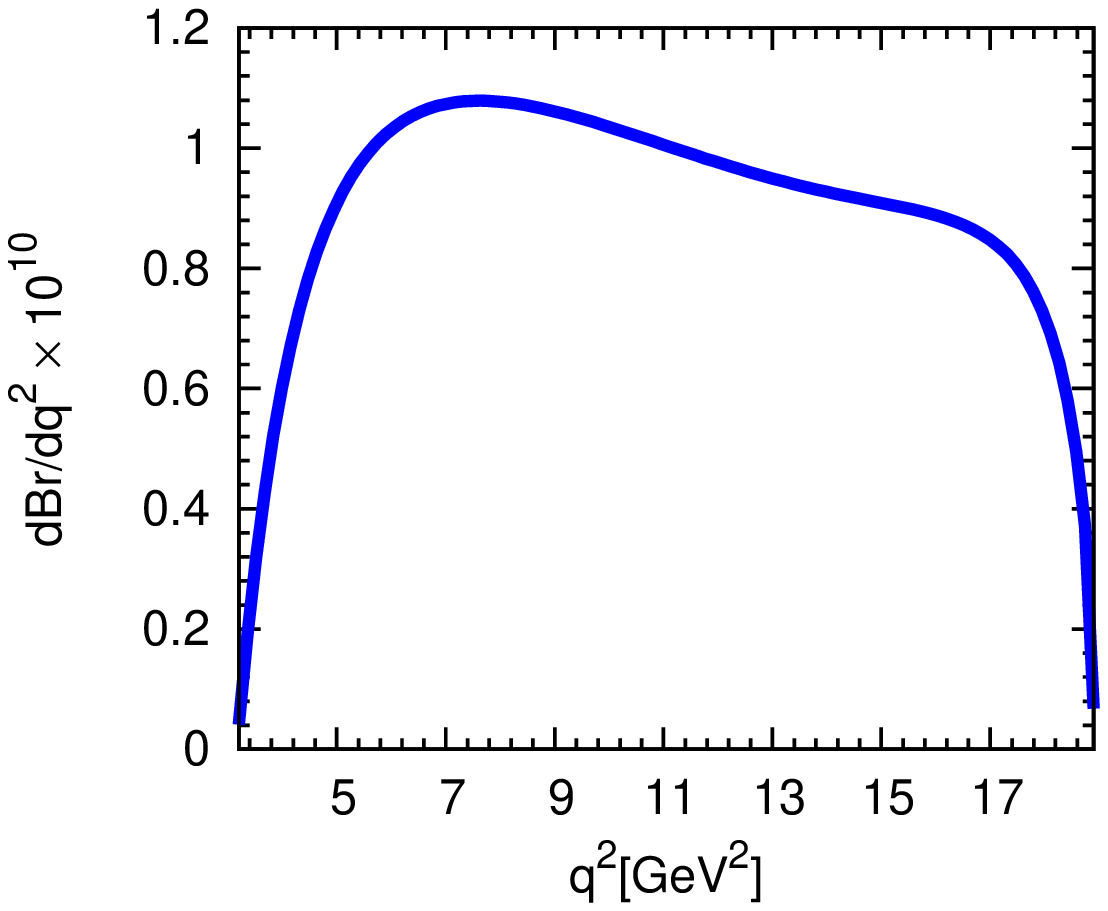}\\
\includegraphics [width=6.5cm,height=5.0cm]{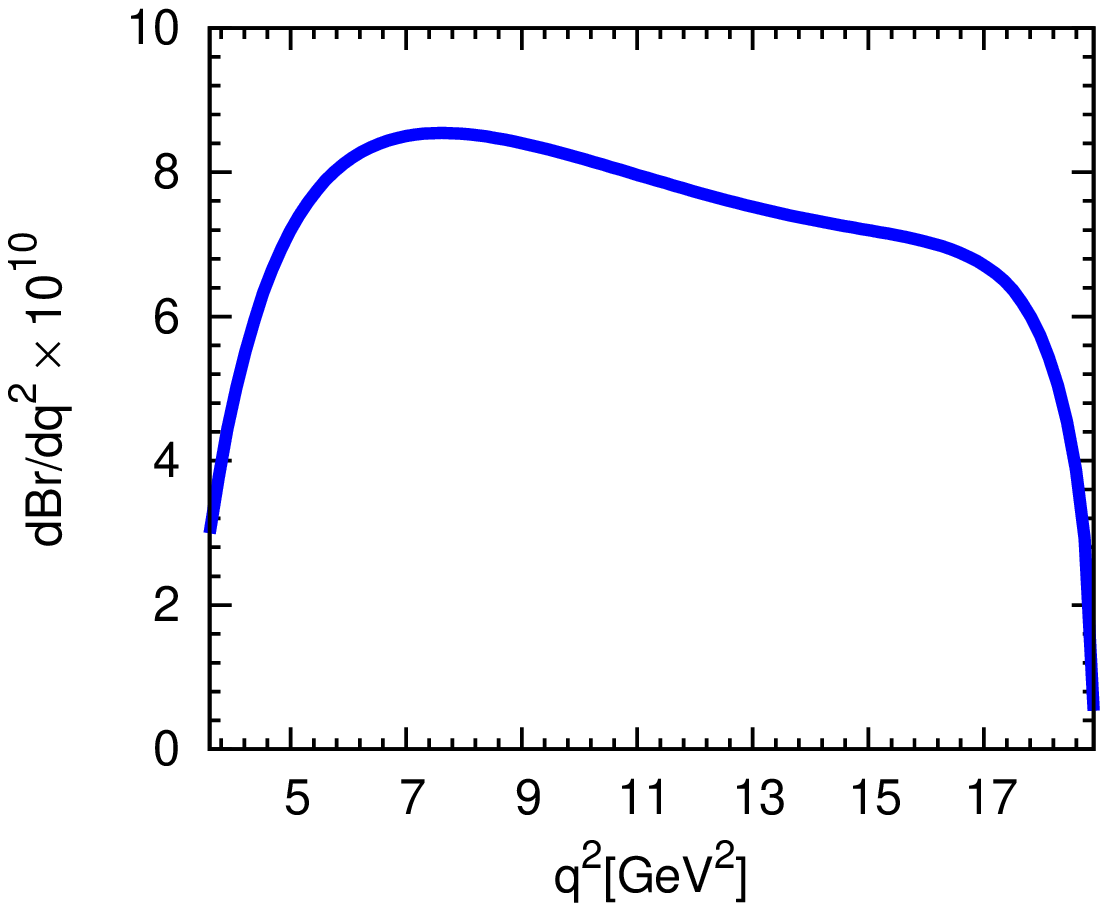}
\caption{The variation of branching ratio of $ B_s \rightarrow \phi \mu^+ e^-$ (left panel), $B_s \rightarrow \phi \tau^+ e^-$ (right panel), and $ B_s \rightarrow \phi \tau^+ \mu^-$ (bottom panel)  with respect to $q^2$ in the scalar leptoquark model.}
\end{figure}
\begin{figure}[h]
\centering
\includegraphics[width=6.0cm,height=5.0cm]{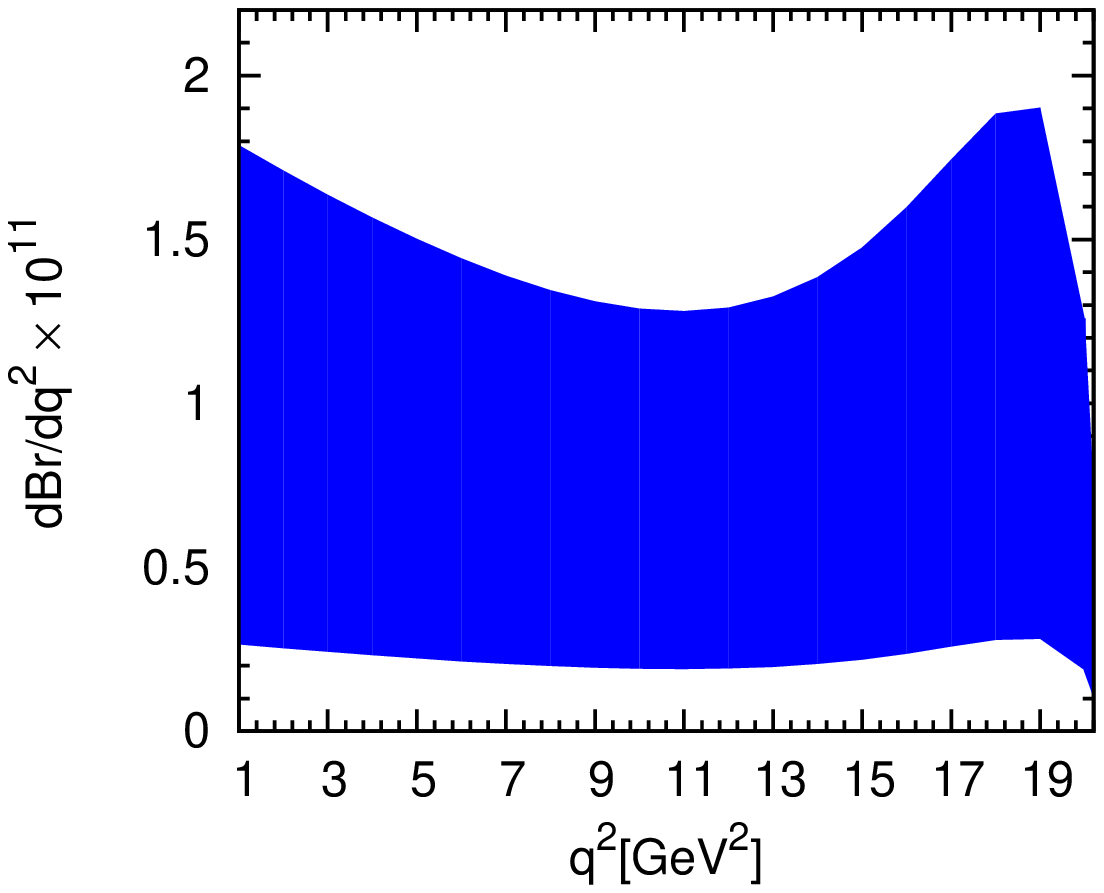}
\includegraphics[width=6.5cm,height=5.0cm]{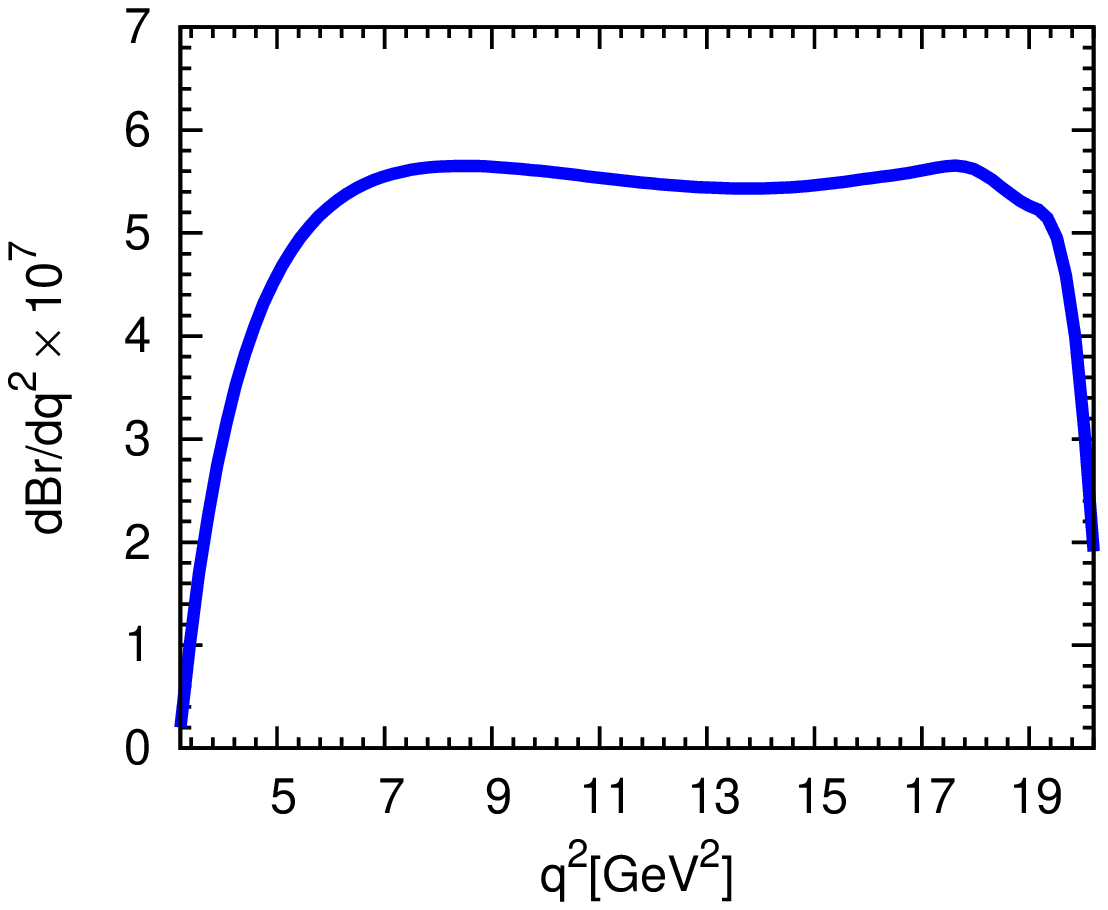}\\
\includegraphics[width=6.5cm,height=5.0cm]{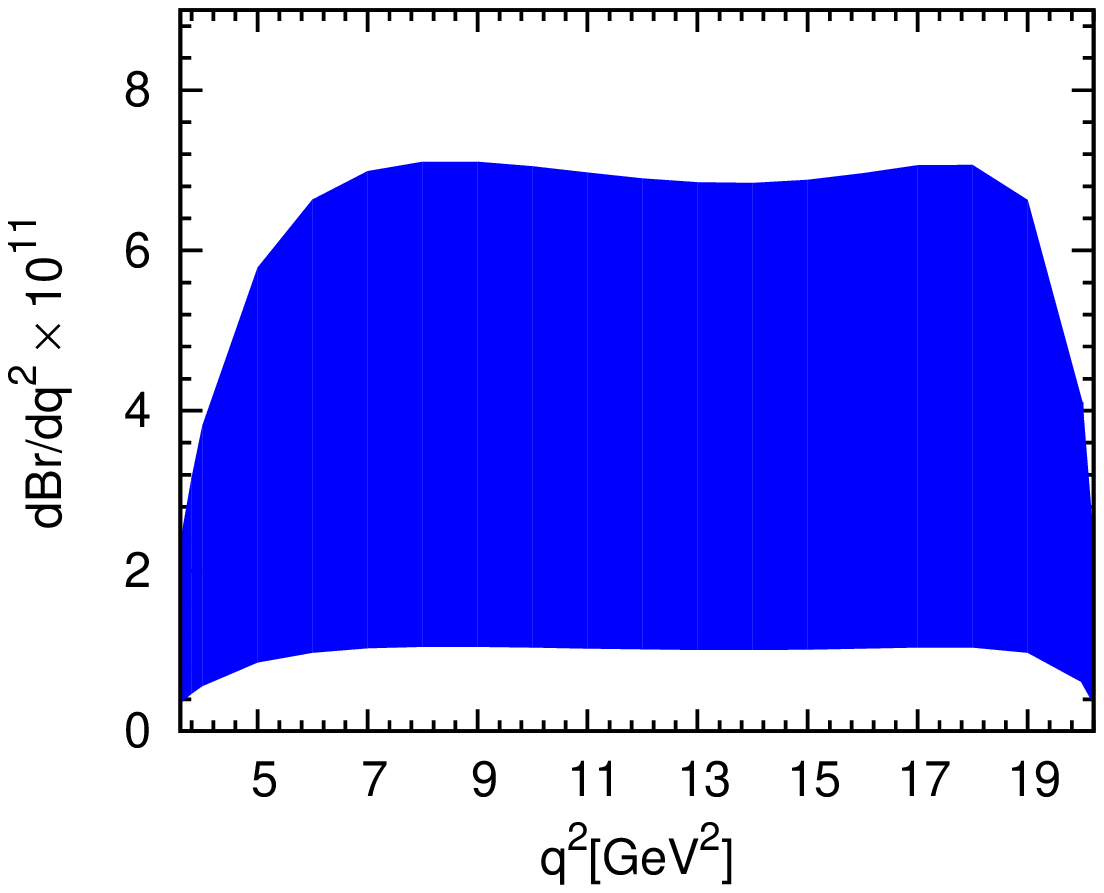}
\caption{The variation of branching ratio of $ B^+ \rightarrow \rho^+ \mu^+ e^-$ (left panel), $B^+ \rightarrow \rho^+ \tau^+ e^-$ (right panel), and $ B^+ \rightarrow \rho^+ \tau^+ \mu^-$ (bottom panel)  with respect to $q^2$ in the scalar leptoquark model.}
\end{figure}

\begin{table}[h]
\caption{The predicted branching ratios for $B_{(s)}^+ \to V^+ (\phi)  l_i^+ l_j^-$  lepton flavour violating decays, where $ V = K^* \rho $ and $l = e, \mu, \tau$. }
\begin{center}
\begin{tabular}{| c | c | c |}
\hline
 Decay process & Predicted BR &  Experimental limit \cite{pdg} \\
 \hline
$B^+ \to K^{* +} \mu^+ e^-$ & $\textless 1.4 \times 10^{-9}$ & $\textless 9.9 \times 10^{-7}$\\
$B^+ \to K^{* +} \tau^+ \mu^-$ & $\textless 1.56  \times 10^{-8}$ &  \ldots \\
$B^+ \to K^{* +} \tau^+ e^-$ & $\textless 2 \times 10^{-9}$ &  \ldots \\

\hline

$B_s \to \phi \mu^+ e^-$ & $\textless 8.2 \times 10^{-10}$ &  \ldots \\
$B_s \to \phi  \tau^+ \mu^-$ & $\textless 1.1 \times 10^{-8}$ &  \ldots \\
$B_s \to \phi \tau^+ e^-$& $ \textless 1.42 \times 10^{-9}$ &  \ldots  \\

\hline

$B^+ \to \rho^+ \mu^+ e^-$ & $ (0.43 - 2.9) \times 10^{-10}$ &  \ldots \\
$B^+ \to \rho^+  \tau^+ \mu^-$ & $ (0.162 - 1.1) \times 10^{-9}$ &  \ldots \\
$B^+ \to \rho^+ \tau^+ e^-$& $\textless 8.73 \times 10^{-6}$ &   \ldots\\

 \hline
\end{tabular}
\end{center}
\end{table}

Recently LHCb has observed  $2.6\sigma$ discrepancy from the SM prediction  in the measurement of the ratio of branching fractions of $B \to K l^+ l^-$  decays into dimuons over dielectrons in the dilepton invariant mass bin $\left( 1 \leq q^2 \leq 6 \right) {\rm GeV^2}$ \cite{lhcb4}.  Analogously, we would like to see whether it would be possible to observe the lepton non-universality effects in semileptonic  LFV decays. We define the ratio of branching ratios of LFV $B^+ \to P^+ l_i^+ l_j^-$ decays to $B^+ \to P^+ \mu^+ e^-$ process as 
\bea
R_{P\mu e}^{l_il_j} = \frac{{\rm BR}\left( B^+ \to P^+ l_i^+ l_j^-\right)}{{\rm BR}\left( B^+ \to P^+ \mu^+ e^-\right)},
\eea
where $l_{i, j}$ stand for all charged leptons. In Fig. 7, we show the $q^2$ variation of $R_{K \mu e}^{\tau e}$  (top-left panel), $R_{K \mu e}^{\tau \mu }$  
(top-right panel), $R_{K^* \mu e}^{\tau e}$  (bottom-left panel) and  $R_{K^* \mu e}^{\tau \mu }$  
(bottom-right panel)  in the $X=(3,2,7/6)$ LQ model. The integrated values of these ratios in the $X=(3,2,7/6)$ leptoquark model for both scalar and vector meson are presented  in Table IV. 
Here we have considered only the  upper limit value  for the LQ couplings. Since the couplings of leptoquarks differ when they couple to different generations of quarks and leptons, no definitive conclusion can be inferred from these results. However, if these ratios will be measured in future, they will provide interesting insight about  the nature of new physics.

In addition, the ratio of branching fractions of $B^+ \to \pi^+ \mu^+ \mu^-$ over  $B^+ \to K^+ \mu^+ \mu^-$ process has been measured by LHCb collaborations \cite{lhcb8} as
\bea
\frac{{\rm BR}(B^+ \to \pi^+ \mu^+ \mu^-)}{{\rm BR}(B^+ \to K^+ \mu^+ \mu^-)}=0.053 \pm 0.014 ({\rm stat}) \pm 0.001~ ({\rm syst}).
\eea
Analogously, one can also define the ratio of branching fractions of $B^+ \to \pi^+ l_i^+ l_j^-$ and $B^+ \to K^+ l_i^+ l_j^-$ LFV processes as 
\bea
R_{\pi K}^{l_i l_j} = \frac{{\rm BR}\left( B^+ \to \pi^+ l_i^+ l_j^-\right)}{{\rm BR}\left( B^+ \to K^+ l_i^+ l_j^-\right)} .
\eea
The variation of $R_{\pi K}^{\mu e}$ (left panel), $R_{\rho K^*}^{\mu e}$ (right panel) and $R_{\rho \phi}^{\mu e}$ (bottom panel) with respect to $q^2$ are shown in Fig. 8.
We present the predicted values of above defined ratios in Table IV, (where we have considered only the  upper limit value  for the LQ couplings), along with the values for the corresponding vector meson case.  The study of the above ratios in the leptoquark model  provides  additional new observables, which could be searched at  LHCb and other $B$-factories.
\begin{figure}[h]
\centering
\includegraphics[width=6.5cm,height=5.0cm]{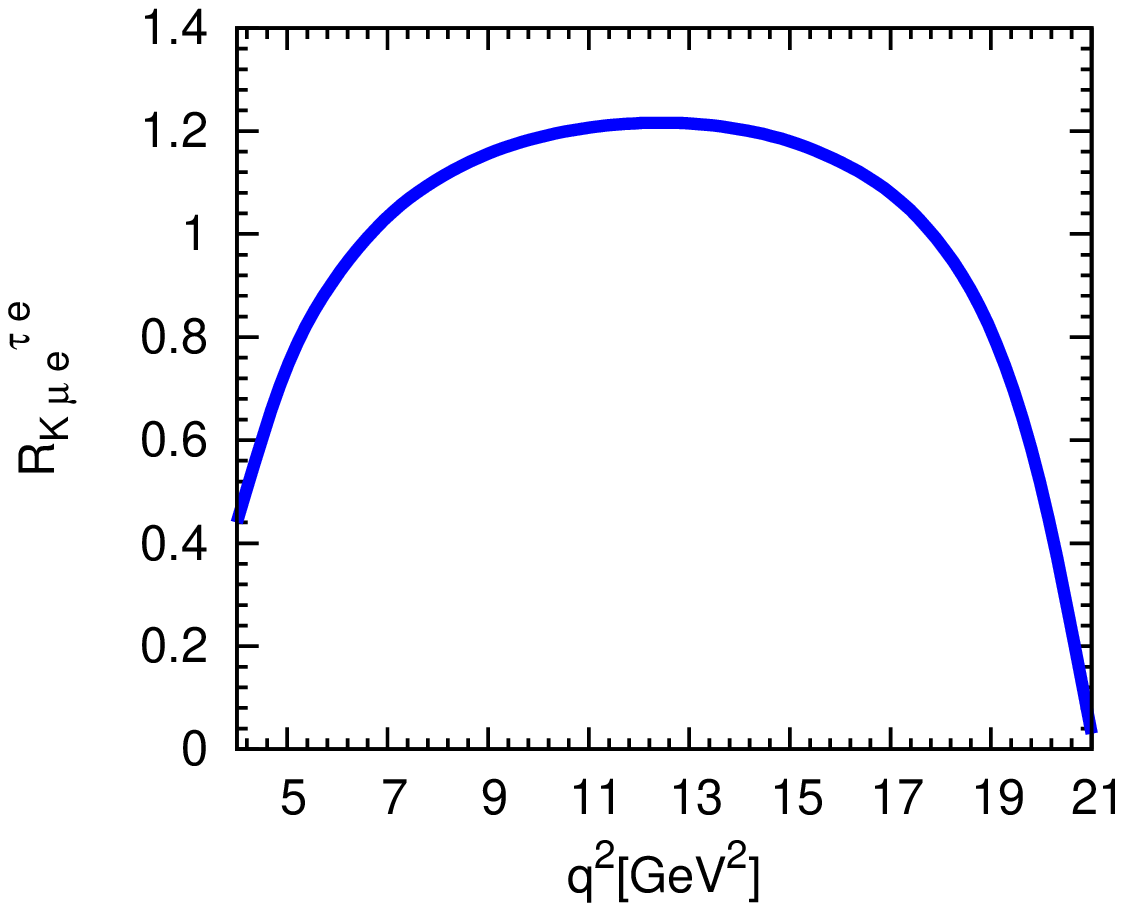}
\includegraphics[width=6.5cm,height=5.0cm]{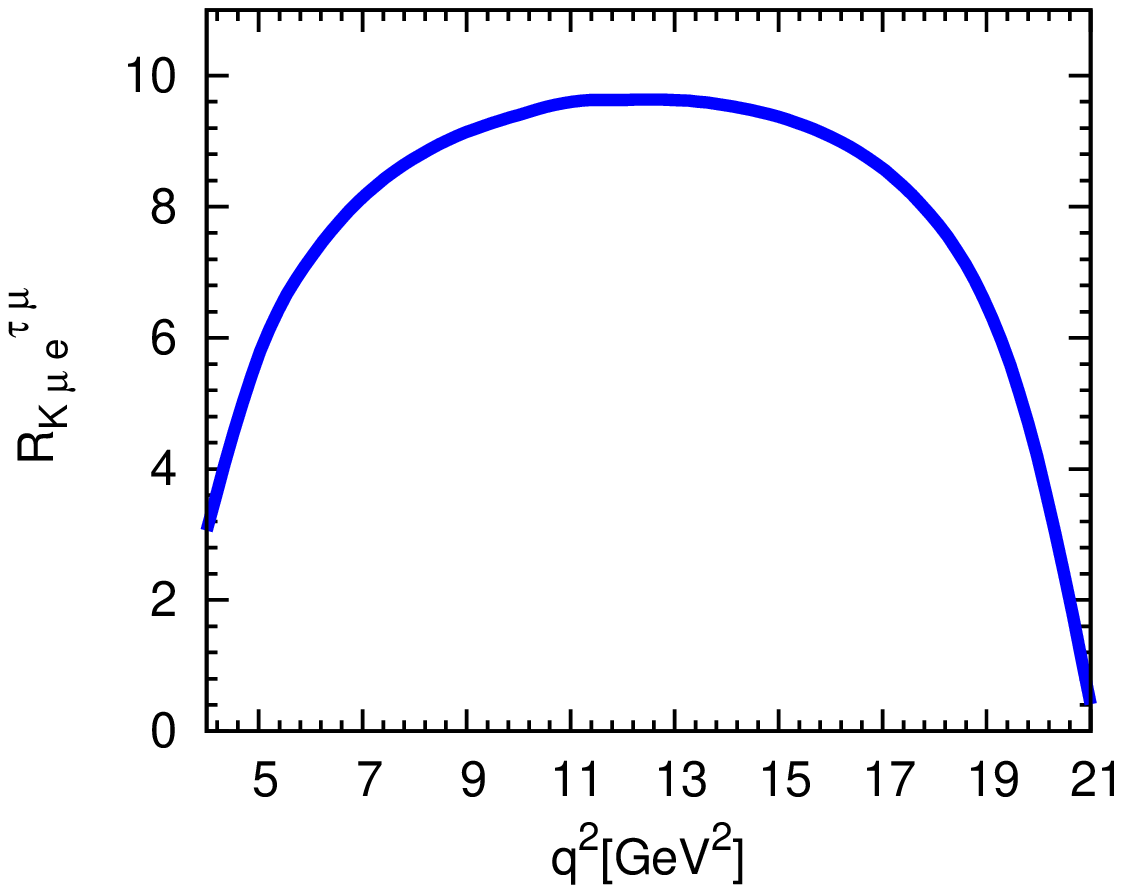}\\
\includegraphics[width=6.5cm,height=5.0cm]{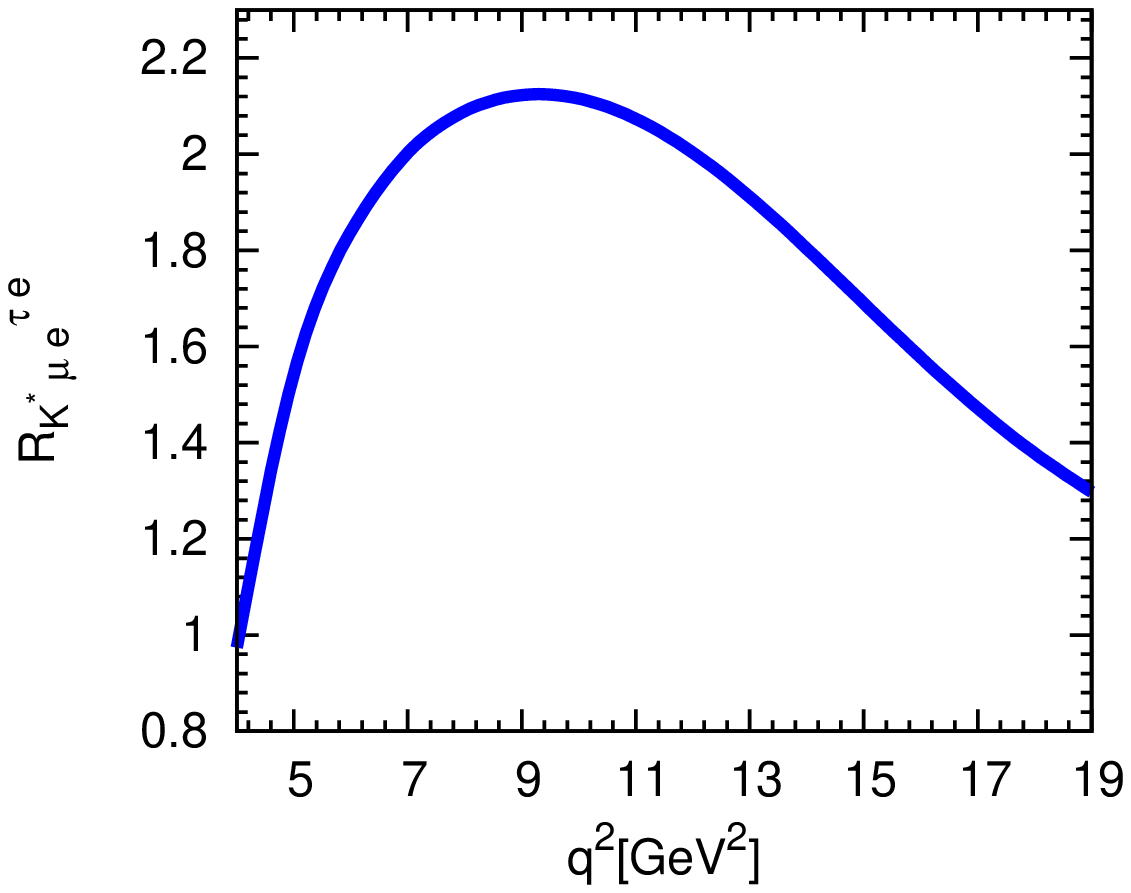}
\includegraphics[width=6.5cm,height=5.0cm]{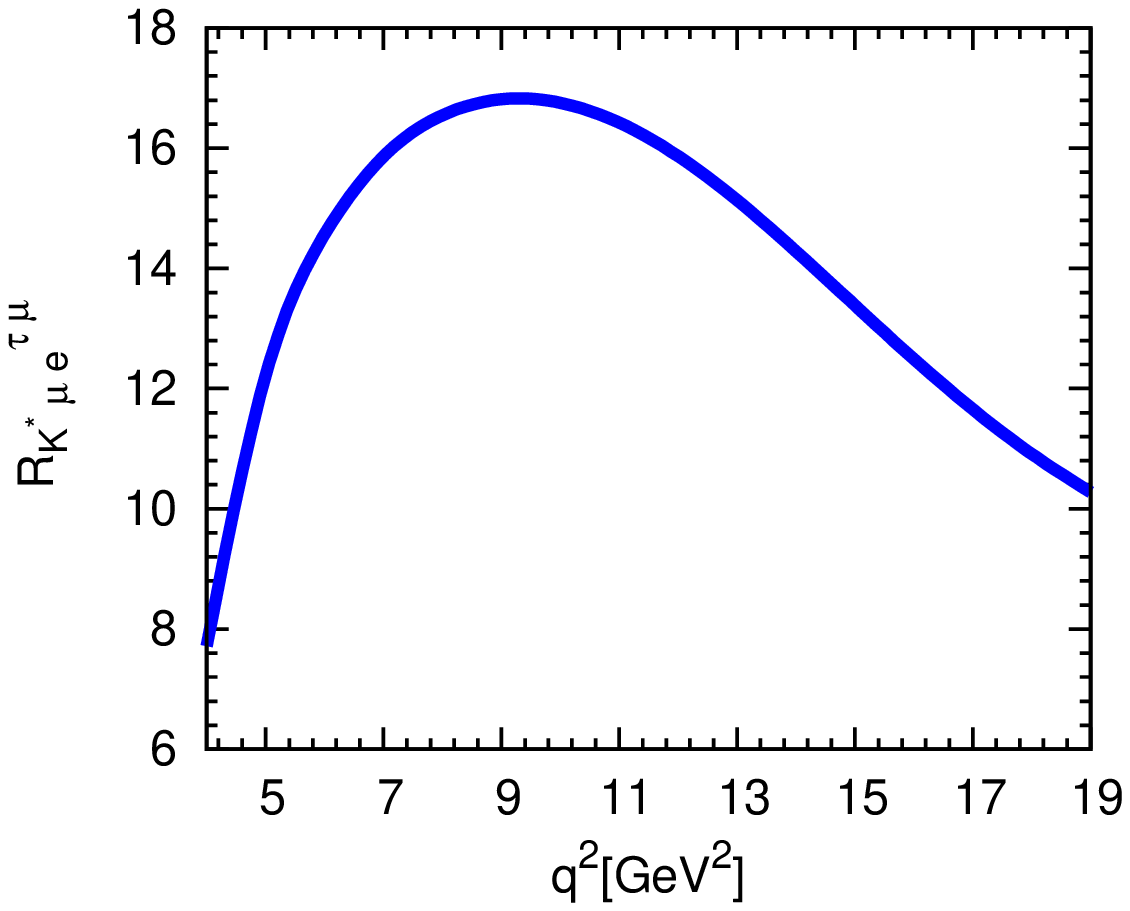}\\
\caption{The variation of observables   $ R_{K\mu e}^{\tau e}$ (top-left panel),  $ R_{K\mu e}^{\tau \mu}$  (top-right panel),
$ R_{K^*\mu e}^{\tau e}$ (bottom-left panel),  $ R_{K^*\mu e}^{\tau \mu}$  (bottom-right panel), with respect to  $q^2$ in the scalar leptoquark model $X(3,2,7/6)$.}
\end{figure}
\begin{figure}[h]
\centering
\includegraphics[width=6.5cm,height=5.0cm]{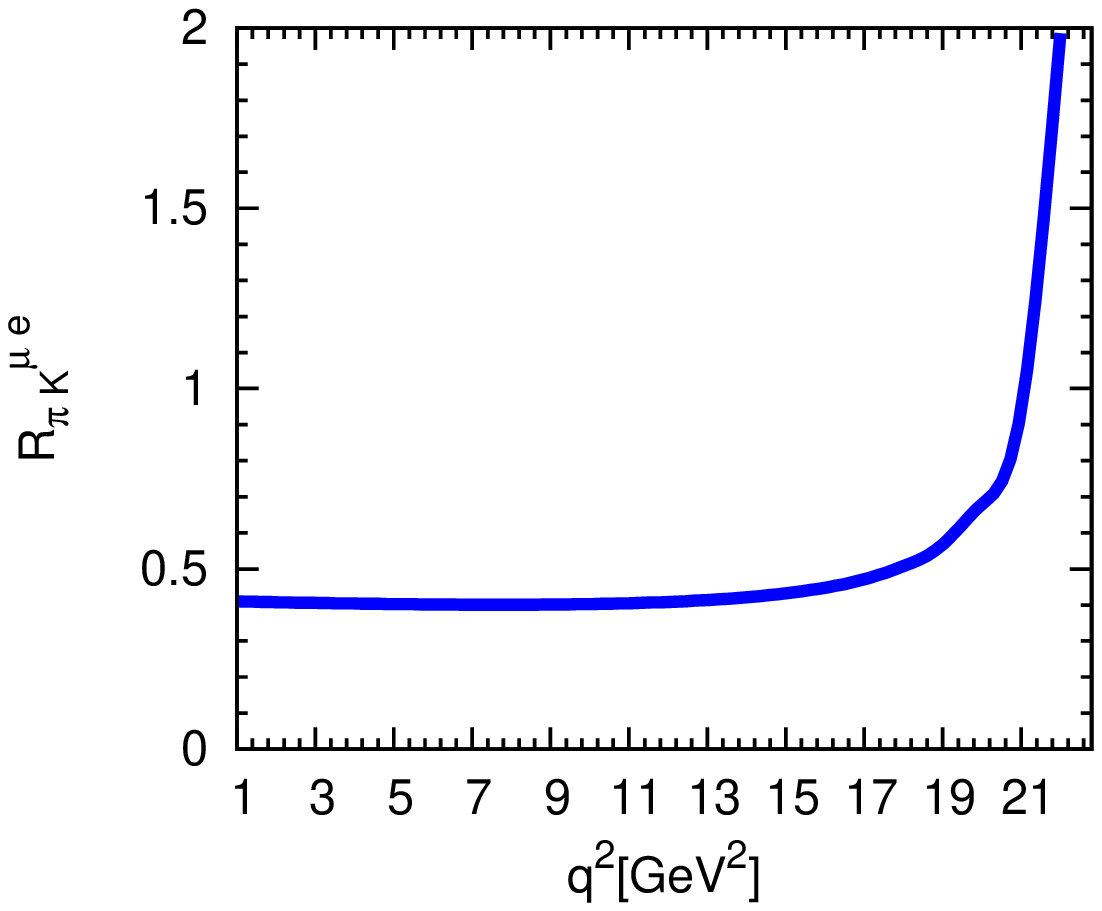}
\includegraphics[width=6.5cm,height=5.0cm]{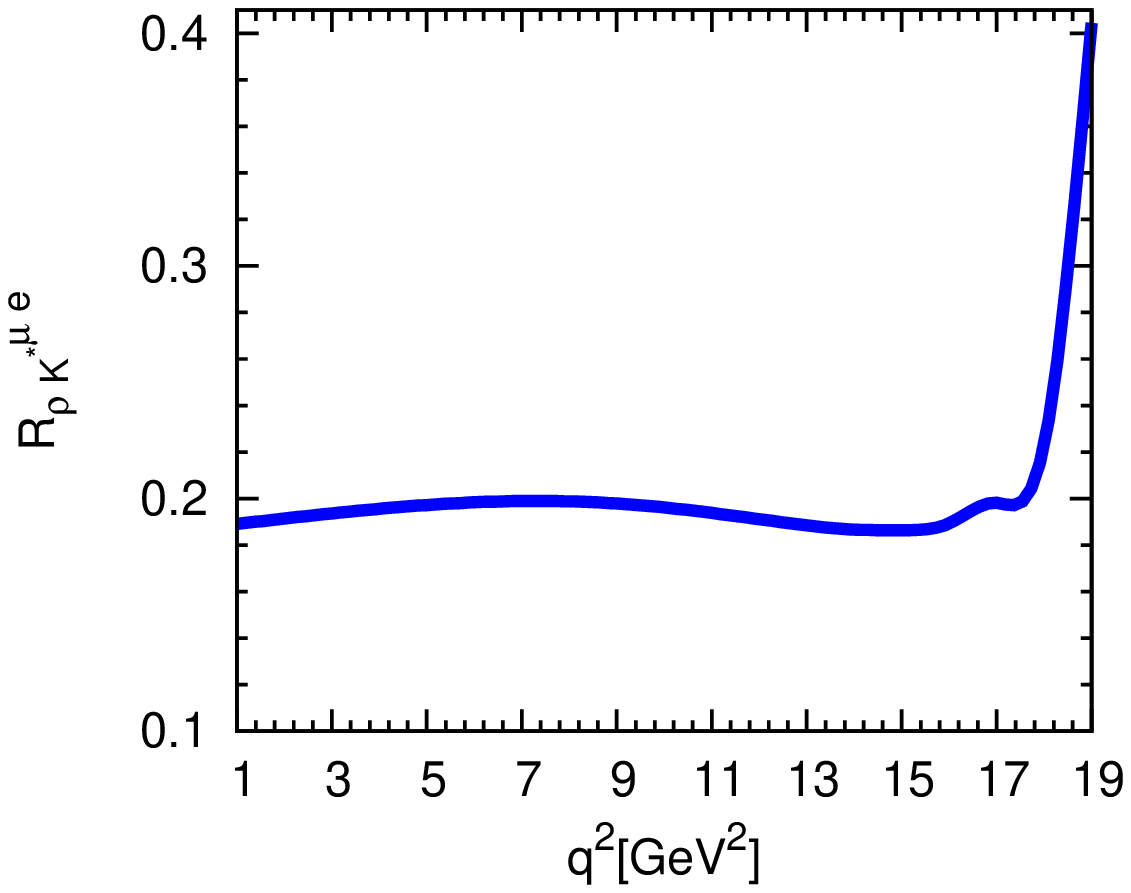}\\
\includegraphics[width=6.5cm,height=5.0cm]{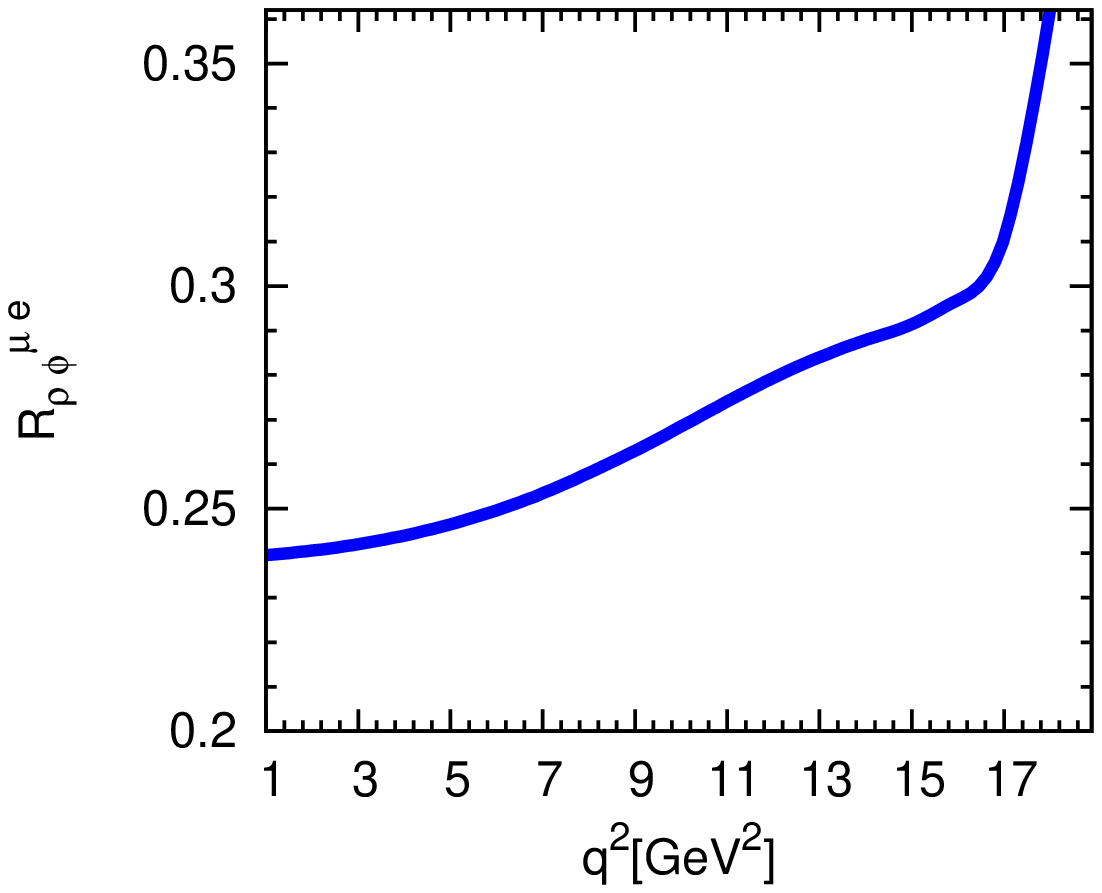}
\caption{The variation of observables  $ R_{\pi K}^{\mu e}$ (left panel), $R_{\rho K^*}^{\mu e}$ (right panel) and $ R_{\rho \phi}^{\mu e}$ (bottom panel) 
with respect to  $q^2$ in the scalar leptoquark model.}
\end{figure}

\begin{table}[h]
\caption{The predicted integrated values of $R_{P(V) \mu e}^{l_i l_j}$ and  $R_{PP'(VV')}^{l_i l_j}$ observables in the $X=(3,2,7/6)$ LQ model, where 
$P,P'=K, \pi$, $ V,V' = K^*, \phi, \rho $ and $l = e, \mu, \tau$. }
\begin{center}
\begin{tabular}{|  c | c | c | c |}
\hline
~~ Observables~~ & ~~Predicted values ~~ & ~~ Observables~~ & ~~Predicted values ~~\\
 \hline
$R_{K \mu e}^{\tau \mu}$ & $6.47 $ & $R_{\pi K}^{\mu e}$ & $0.453 $ \\
$R_{K \mu e}^{\tau e}$ & $0.82 $  
  &   $R_{\pi K}^{\tau \mu}$ & $0.14$ \\
$R_{K^* \mu e}^{\tau e}$ & $ 1.43$ & $R_{\rho K^*}^{\tau \mu}$ & $0.07$   \\
$R_{\pi \mu e}^{\tau \mu}$ & $1.95$ & $R_{\rho K^*}^{\mu e}$ & $0.21$ \\
 $R_{\rho \mu e}^{\tau \mu}$ & $ 3.79$
& $R_{\rho \phi}^{ \mu e}$ & $0.354$ \\

 $R_{\phi \mu e}^{\tau e}$ & $ 1.73$ & $R_{\rho \phi}^{ \tau \mu}$& $0.1$  \\
 \hline
\end{tabular}
\end{center}
\end{table}

\section{$K_{L, S} \to \mu^+ \mu^- \left( e^+ e^-\right)$}
In this section, we study the rare leptonic decays of $K$ meson and  would like to see how the scalar leptoquarks affect these processes. 
The rare $K_L \to \mu^+ \mu^-$ decay is CP conserving and provides valuable information on the short distance physics of  $|\Delta S=1|$ FCNC transitions.   
This decay mode acquires dominant contributions from the long distance two photon intermediates state $K_L \to \gamma^* \gamma^* \to \mu^+ \mu^-$. 
Therefore, although its branching ratio is measured precisely with value ${\rm Br}(K_L \to \mu^+ \mu^-)=(6.84 \pm 0.11) \times 10^{-9}$ \cite{pdg},
the SM prediction is not reliable because of the long-distance effects. However, the dispersive method gives the estimate of the short-distance part as
 ${\rm Br}(K_L \to \mu^+ \mu^-)|_{SD}< 2.5  \times 10^{-9}$ \cite{Kaon-SM-BR}. 
The short distance (SD) part can be calculated reliably and is on the same footing as 
$K^+ \to \pi^+ \nu \bar{\nu}$ process except the difference of lepton line in the box diagram. 
 The SD contribution to  the effective Hamiltonian for $K_L \to \mu^+ \mu^-$ process in the SM is given by \cite{buras}
\bea
\mathcal{H}_{eff} = \frac{G_F}{\sqrt{2}} \frac{\alpha}{2 \pi \sin^2\theta_W} \left(\lambda_c Y_{NL} + \lambda_t Y(x_t) \right)  \left( \bar{s} \gamma^\mu (1-\gamma_5)d \right) \left(\bar{\mu} \gamma_\mu (1-\gamma_5)\mu \right),
\eea
where the functions $Y_{NL}$ and $Y(x_t)$ are the contributions from charm and top quark respectively and the $Y(x_t)$ function 
in the next-to-leading order (NLO) is given as \cite{buras2}
\bea
Y(x_t) = \eta_Y \frac{x_t}{8} \Bigg( \frac{4-x_t}{1-x_t} + \frac{3x_t}{\left(1-x_t \right)^2} \ln x_t \Bigg).
\eea
The branching ratio  for the  SD part in the SM is 
\bea
{\rm BR}(K_L \to \mu^+ \mu^-)|_{\rm SD} = \frac{G_F^2 }{2\pi }\tau_{K_L} |V_{us}^* V_{ud}|^2\Big(\frac{m_\mu}{M_K}\Big)^2 
\sqrt{1-\frac{4m_\mu^2}{M_K^2}} f_K^2 M_K^3 \left|\Delta^K_{SM} \right|^2,
\eea
where 
\bea
\Delta_{SM}^K = \frac{\alpha \left( \lambda_c Y_{NL} + \lambda_t Y(x_t) \right)}{2\pi \sin^2\theta_W V_{us}^* V_{ud}}.
\eea
Now using the interaction Lagrangian  (\ref{lq7-lagrangian}) for $X=(3,2,7/6)$ scalar leptoquark  the relevant Hamiltonian of $K_{L,S} \to \mu^+ \mu^-$ process in the LQ model is given as
\bea
\mathcal{H}_{LQ}=\frac{\lambda_\mu^{21} {\lambda_\mu^{22}}^*}{8M_Y^2} \left(\bar{s}\gamma^\mu \left(1-\gamma_5\right)d\right) \left(\bar{\mu}\gamma^\mu \left(1-\gamma_5\right)\mu\right). \label{Kaon-LQ_HN}
\eea
 Here we consider $K_L$ as a pure CP-odd state and $K_S$ as CP-even, which are decaying into CP-odd $\mu^+ \mu^-$ final state. 
If we include the contributions from both the $K^0$ and $\bar{K^0}$ decay amplitude,  the leptoquark couplings given in Eq. (\ref{Kaon-LQ_HN}) 
will be replaced by $\sqrt{2}  \frac{ {\rm Re} (\lambda_\mu^{21} {\lambda_\mu^{22}}^*)}{M_Y^2}$ for $K_L \to \mu^+ \mu^-$ decay and 
for $K_S \to \mu^+ \mu^-$ process the coupling will be $ \sqrt{2} \frac{ {\rm Im} (\lambda_\mu^{21} {\lambda_\mu^{22}}^*)}{M_Y^2} $.  
The decay processes  $K_{L,S} \to \mu^+ \mu^-$ are studied  in the leptoquark model  in Refs. \cite{dorsner, Girish}.
Including the leptoquark contribution, the branching ratios of $K_L \to \mu^+ \mu^-$ process is given by
\bea
{\rm BR}(K_L \to \mu^+ \mu^-)=&&\frac{f_K^2 m_\mu^2 M_K}{8\pi} \tau_{K_L} \sqrt{1-\frac{4m_\mu^2}{M_K^2}} \nn \\ &&\times ~ \Bigg | \frac{G_F}{\sqrt{2}} \frac{\alpha}{2 \pi \sin^2\theta_W} \left(\lambda_c Y_{NL} + \lambda_t Y(x_t) \right) +  \sqrt{2} \frac{ {\rm Re} (\lambda_\mu^{21} {\lambda_\mu^{22}}^*)}{8M_Y^2}  \Bigg |^2
\eea
and the value of the predicted branching ratio  is presented in Table V. The masses of different particles and the life times of $K_{L, S}$ mesons 
have been taken from \cite{pdg} and we use the same scaling law i.e. $\lambda^{ij} \simeq (m_i/m_j)^{1/4}\lambda^{ii}$ with $j \textgreater i$ 
to obtain the bounds on the required leptoquark couplings  for $K_{L, S} \to \mu^+ \mu^- (e^+ e^-)$ transitions.  Similarly we calculate 
the branching ratios of $K_{S} \to \mu^+ \mu^-$ and $K_{L, S} \to e^+ e^-$ processes and the corresponding values are listed in Table V. 
For the $K_L \to e^+e^-$ mode, the experimental measurement ${\rm Br}(K_L \to e^+e^-) =(9_{-4}^{+6}) \times 10^{-12}$  \cite{pdg} is in good agreement with
the SM long-distance estimate  ${\rm Br}(K_L \to e^+e^-)|_{LD} =(9\pm0.5) \times 10^{-12}$  \cite{valencia}, hence, the short distance contribution
is almost negligible.
The SM prediction for  $K_{S} \to \mu^+ \mu^- ~(e^+ e^-)$ process is  $2 \times 10^{-6} ~(8 \times 10^{-9} ) \times {\rm Br}(B_S \to \gamma \gamma) 
\sim 10^{-11}~ (10^{-14})$ respectively \cite{KS-SM}. 

\begin{table}[h]
\caption{The predicted branching ratios for $K_{L, S} \to \mu^+ \mu^-(e^+e^-) $ processes (short-distance part) in the $X=(3,2,7/6)$ LQ model. }
\begin{center}
\begin{tabular}{| c | c | c |}
\hline
 Decay processes & Predicted BR  &  Experimental values \cite{pdg} \\
 \hline
$K_L \to \mu^+ \mu^-$ & $(0.13-2.2) \times 10^{-9} $ &  $\left(6.84 \pm 0.11 \right) \times 10^{-9} $ \\
$K_L \to e^+ e^-$ & $1.1 \times 10^{-12} $ & $9^{+6}_{-4} \times 10^{-12} $ \\
$K_S \to \mu^+ \mu^-$ & $\textless ~ 2.23 \times 10^{-11} $ & $\textless 9 \times 10^{-9} $ \\
$K_S \to e^+ e^-$ & $\textless~ 1.9 \times 10^{-15} $ & $\textless 9 \times 10^{-9}$ \\
 \hline
\end{tabular}
\end{center}
\end{table}

\section{$K_{L} \to \mu^\mp e^\pm$}
Next, we would like to  investigate the rare leptonic LFV $K_{L} \to \mu^\mp e^\pm$ decays.
The effective Hamiltonian for $K_{L} \to \mu^+ e^-$ LFV decays in the $X=(3,2,7/6)$ scalar leptoquark model is 
\bea
\mathcal{H}_{LQ}=\frac{ \lambda_\mu^{22} {\lambda_\mu^{11}}^* }{8M_Y^2} \left(\bar{s}\gamma^\mu \left(1-\gamma_5\right)d\right) \left(\bar{\mu}\gamma^\mu \left(1-\gamma_5\right)e\right),
\eea
and for  $K_{L} \to \mu^- e^+$ process
\bea
\mathcal{H}_{LQ}=\frac{ \lambda_\mu^{12} {\lambda_\mu^{21}}^* }{8M_Y^2} \left(\bar{s}\gamma^\mu \left(1-\gamma_5\right)d\right) \left(\bar{e}\gamma^\mu \left(1-\gamma_5\right)\mu \right). 
\eea
In the literature \cite{lfv kaon, Girish} the LFV decay of kaon has been studied in the leptoquark and other new physics model.
The corresponding branching ratio in the leptoquark model is given by
\bea
{\rm BR}(K_L \to \mu^\mp e^\pm)&=&\frac{f_K^2 \tau_{K_L}}{512 \pi M_K^3} \Bigg | \frac{ \lambda_\mu^{22} {\lambda_\mu^{11}}^*+\lambda_\mu^{12} {\lambda_\mu^{21}}^* }{M_Y^2} \Bigg |^2 \nn \\ & \times & \sqrt{\left(M_K^2-m_\mu^2- m_e^2 \right)^2 - 4 m_\mu^2 m_e^2} ~ \Big(M_K^2\left(m_\mu^2 + m_e^2\right)-\left(m_\mu^2 - m_e^2\right)^2 \Big). ~~~~
\eea
Now using the particle masses from \cite{pdg} and the scaling ansatz for LQ couplings, the predicted branching ratios of $K_{L} \to \mu^\mp e^\pm$ process is
\bea
{\rm BR}(K_L \to \mu^\mp e^\pm)&=& 7.17 \times 10^{-13}. 
\eea
There exists only the upper limit on branching ratio of $K_L \to \mu^\mp e^\pm$ decay with value ${\rm BR}(K_L \to \mu^\mp e^\pm) 
\textless 4.7 \times 10^{-12}$ at 90\% C.L. \cite{pdg} and our predicted result is within the experimental limit.
   
\section{conclusion}
In this paper, we have studied the rare lepton flavour violating semileptonic $B$ meson decays  in the scalar leptoquark model. These decays are extremely rare in the SM as they occur at loop level.  They  are  further suppressed due to the tiny neutrino masses in one of the loop.  However, in the scalar leptoquark model, these decays can occur at the tree level as the  leptoquark couples to quark and lepton simultaneously thereby mediating the LFV processes at tree level.  The scalar leptoquarks which  do not have baryon number violation in the perturbation theory forbid proton decay  and could be light enough to
be accessible in accelerator searches. There are only two such leptoquarks $X(3,2,7/6)$ and $X(3,2,1/6)$ which could satisfy these conditions. We considered such leptoquarks and studied the various lepton flavour violating decays. The  leptoquark parameter space is constrained using the recently measured branching ratios of $B_q \rightarrow l^+ l^-$ 
from LHCb and CMS experiments and  using such constrained parameters we estimated the  branching ratios of LFV decays such as $B^+ \to K^+ (\pi^+) l_i^+ l_j^-$,  
$B^+ \to (K^{* +}, \rho^+ ) l_i^+ l_j^-$ and $B_s \to \phi l_i^+ l_j^-$. We study the ratios of various combination of LFV decays in order to check the presence of lepton non-universality.  We also predicted the branching ratios of leptonic Kaon decays $(K_{L, S} \to \mu^+ \mu^-)$ and the LFV $K_{L} \to \mu^+ e^-$ processes in the leptoquark model.
We found that our predicted values are  within the present  experimental limits,  the observation of which in the LHCb or upcoming Belle II experiments would provide unambiguous signal  of new physics.\\


{\bf Acknowledgments}

We would like to thank Science and Engineering Research Board (SERB),
Government of India for financial support through grant No. SB/S2/HEP-017/2013.


\end{document}